\documentclass[nofootinbib, twocolumn, aps, prd]{revtex4-1}

\usepackage{graphicx}
\usepackage{natbib}
\usepackage{color}

\newcommand{\beq}{\begin{equation}}
\newcommand{\beqa}{\begin{eqnarray}}
\newcommand{\eeq}{\end{equation}}
\newcommand{\eeqa}{\end{eqnarray}}

\newcommand{\simgt}{\lower.5ex\hbox{$\; \buildrel > \over \sim \;$}}
\newcommand{\simlt}{\lower.5ex\hbox{$\; \buildrel < \over \sim \;$}}

\newcommand{\bd}[1]{\mbox{\boldmath $#1$}}

\usepackage[usenames,dvipsnames]{xcolor}

\begin{document}
\title{
Cross-Correlation of the
Extragalactic Gamma-ray Background \\
with
Luminous Red Galaxies 
}

\author{Masato Shirasaki}
\affiliation{
National Astronomical Observatory of Japan, 
Mitaka, Tokyo 181-8588, Japan}
\email{masato.shirasaki@nao.ac.jp}

\author{Shunsaku Horiuchi}
\affiliation{
Center for Neutrino Physics, Department of Physics, Virginia Tech, Blacksburg, Virginia
24061, USA
}
\email{horiuchi@vt.edu}

\author{Naoki Yoshida}
\affiliation{
Department of Physics, University of Tokyo, Tokyo 113-0033, Japan\\
Kavli Institute for the Physics and Mathematics of the Universe (WPI),
University of Tokyo, Kashiwa, Chiba 277-8583, Japan
}
\affiliation{
CREST, Japan Science and Technology Agency, 4-1-8 Honcho, Kawaguchi, Saitama, 332-0012, Japan
}
\email{naoki.yoshida@ipmu.jp}

\begin{abstract}
  Measurements of the cross-correlation between the extragalactic gamma-ray background (EGB)
  and large-scale structure provide a novel probe of dark matter on extragalactic scales.
  We focus on luminous red galaxies (LRGs) as optimal targets to
  search for the signal of dark matter annihilation.
  We measure the cross-correlation function of the EGB taken from the Fermi Large Area Telescope
  with the LRGs from the Sloan Digital Sky Survey. Statistical errors are calculated
  using a large set of realistic mock LRG catalogs.
  The amplitude of the measured cross-correlation is consistent with null detection.
  Based on an accurate theoretical model of the distribution of dark matter associated
  with LRGs,
  we exclude dark matter annihilation cross-sections over 
  $\langle \sigma v\rangle =3\times10^{-25}-10^{-26}\, {\rm cm}^3 \,{\rm s}^{-1}$
  for a 10 GeV dark matter.
  We further investigate systematic effects due to uncertainties in the Galactic gamma-ray
  foreground emission, which we find to be an order of magnitude smaller
  than the current statistical uncertainty. We also estimate the contamination from 
  astrophysical sources in the LRGs by using
  known scaling relations between gamma-ray luminosity and star-formation rate,
  finding them to be negligibly small.
  Based on these results, we suggest that LRGs remain ideal targets for probing dark
  matter annihilation with future EGB measurement and galaxy surveys.
Increasing the number of LRGs in upcoming galaxy surveys such as LSST would 
lead to big improvements of factors of several in sensitivity. 
\end{abstract}

\maketitle

\section{\label{sec:intro}INTRODUCTION}

Dark matter (DM) is invoked to 
explain a broad range of astronomical and cosmological observations.
DM constitutes $\sim$85\% of the 
matter content of the Universe, and thus plays an essential role in the development
of the rich structure of the Universe, such as galaxies and clusters of galaxies.
Although the nature of DM still remains unknown,
weakly-interacting massive particles (WIMPs) are among the most attractive 
particle physics candidates.
In the WIMP hypothesis, some unknown weakly-interacting particle (DM) is initially in thermal
equilibrium in the early Universe.
The time evolution of the mean number density of the particles is
determined by the DM annihilation rate and the expansion rate of the Universe.
Interestingly, the observed abundance of DM
can be naturally explained by WIMPs with mass in the range of 10 GeV -- 10 TeV 
if their annihilation cross-section is of the order the cross-section 
for weak interactions \cite{1996PhR...267..195J}.

Several ``local" gamma-ray measurements have already
probed interesting parameter ranges of annihilating DM.
Such probes rely on searching for gamma rays from nearby regions where the DM
density is expected to be large, since 
the production rate of the gamma rays is proportional to the DM density squared.
Thus, Milky Way satellite galaxies \cite{Ackermann:2013yva, 2015PhRvD..91j2001B, Ackermann:2015zua}  
or the Galactic center \cite{Abazajian:2012pn, Abazajian:2014fta} are considered to 
be the most promising targets for searching for the signal of DM annihilation.

Recently, the cross-correlation of the extragalactic gamma-ray background (EGB) 
with large-scale structure (LSS) has been proposed 
as a complementary probe 
\cite{Camera:2012cj, 2014FrP.....2....6F, 2014PhRvD..90b3514A, Ando:2014aoa, 2015JCAP...06..029C}.
In addition to known gamma-ray sources such as blazers, 
misaligned AGNs, and star-forming galaxies, 
the DM distributed on extragalactic scales should contribute to the observed EGB. 
Although the mean EGB intensity can be explained 
by unresolved astrophysical sources \cite{2015ApJ...800L..27A},
the anisotropy in the diffuse gamma-ray sky contains, in principle, rich information about
any unresolved DM source contributions
(e.g., see Ref~\cite{Fornasa:2015qua} for review). 
This is because the DM density distribution in the 
Universe is expected to be highly inhomogeneous due to the nonlinear gravitational growth of
structure. Therefore, any observational tracers of the DM distribution should correlate with 
the extragalactic gamma-ray map. In fact, the latest gamma-ray data taken from the Fermi 
Large Area Telescope (LAT) have already shown a positive correlation with several known 
tracers of DM distribution such as nearby galaxies, radio galaxies, and quasars \cite{2015ApJS..217...15X},
as well as the reconstructed mass distribution from weak lensing effect
on the cosmic microwave background \cite{2015ApJ...802L...1F}.

Given the successful detections of the cross-correlation 
between the EGB and LSS \cite{2015ApJS..217...15X, 2015ApJ...802L...1F},
it is important and timely to identify optimal targets to 
possibly {\it detect} the signal of DM annihilation on a statistical basis.
Low-redshift galaxies are among the most promising targets for cross-correlation, 
but it is thought to be difficult to disentangle the contributions from
astrophysical sources and DM annihilation 
\cite{Cuoco:2015rfa}.
In order to break the degeneracy, 
Refs~\cite{Ando:2014aoa, 2015JCAP...06..029C} proposed a tomographic approach, 
i.e., measuring the cross-correlation in separate bins in redshift and/or energy.

In the present paper, we propose an alternative optimal target to search 
for DM annihilation through cross-correlation studies.
We consider the following three conditions:
(i) the targets must have accurate redshift information so that a volume limited sample
    can be generated,
(ii) any intrinsic gamma-ray luminosity from astrophysical sources should be small,
and 
(iii) the statistical properties of the host DM can be determined observationally.
Among the possible candidates,
we focus on the luminous red galaxies (LRGs) selected from the Sloan Digital Sky Survey (SDSS).
LRGs have the advantages of having accurate spectroscopic redshifts, low star-formation rates, 
and a well-constrained relation of their underlying DM distribution through clustering measurements 
and weak lensing analyses.
We thus measure the cross-correlation of the EGB with the LRGs to search for 
DM annihilation signatures. We calculate statistical errors using realistic mock LRG 
catalogs, and study in detail possible systematic effects on the cross-correlation measurement.
In particular, we estimate the systematic error 
due to the uncertainty in the Galactic gamma-ray foreground modeling and star-forming activity in the LRGs.

The paper is organized as follows.
In Section~\ref{sec:DMann}, we describe
the contribution to the EGB from DM annihilation.
We also present a theoretical model of cross-correlation of the EGB with LRGs. 
In Section~\ref{sec:data}, we describe the target galaxies and the gamma-ray data. 
The details of the cross-correlation analysis are provided in Section~\ref{sec:cross}.
In Section~\ref{sec:res}, we show the result of our cross-correlation analysis,
and derive constraints on the DM annihilation cross-section.
Finally, we investigate systematic uncertainties of our measurement in 
Section~\ref{sec:sys}.
Concluding remarks and discussions are given in Section~\ref{sec:con}. 
Throughout the paper, we use the standard cosmological parameters
$H_0=100 h \, {\rm km \, s^{-1}}$
with $h=0.7$, $\Omega_{\rm m0}=0.279$, and $\Omega_{\Lambda}=0.721$.

\section{\label{sec:DMann}Dark Matter Annihilation}

The contribution of DM annihilation to the EGB intensity $I_\gamma$ (the number 
of photons per unit energy, area, time, and solid angle) is given by,
\beqa
E_\gamma I_\gamma  
&=&
\frac{c}{4\pi} \int \frac{{\rm d}z}{H(z)(1+z)^4} \,
E^{\prime}_\gamma \frac{{\rm d}N_\gamma}{{\rm d}E^{\prime}_\gamma}
\frac{\langle \sigma v \rangle}{2} 
\nonumber \\
&& \,\,\,\,\,\,\,\, \,\,\,\,\,\,\,\, \,\,\,\,\,\,\,\, \times
\left[ \frac{\rho_{\rm dm}({\bd x}|z)}{m_{\rm dm}} \right]^2
\, e^{-\tau(E^{\prime}_{\gamma},z)},
\label{eq:Intensity}
\eeqa
where $E_\gamma$ is the observed gamma-ray energy, $E^{\prime}_\gamma = (1+z) E_\gamma$
is the energy of the gamma ray at redshift $z$, 
$H(z) = H_0 [\Omega_{\rm m0}(1+z)^3+\Omega_\Lambda]^{1/2}$ is the 
Hubble parameter in a flat Universe, and the exponential factor
takes into account the effect of gamma-ray attenuation during propagation 
owing to pair creation on diffuse extragalactic photons. 
We adopt the gamma-ray optical depth $\tau\left(E'_\gamma, z \right)$ from Ref.~\citep{Gilmore:2011ks}.
The physics of DM is contained in 
 ${\rm d}N_\gamma /{\rm d}E_\gamma$, the gamma-ray spectrum per annihilation;
$\langle \sigma v \rangle$, the annihilation cross-section times the relative velocity averaged 
with the velocity distribution function; $\rho_{\rm dm}({\bd x}|z)$, the DM mass density distribution
at redshift $z$ as a function of spatial coordinate ${\bd x}$; and $m_{\rm dm}$, the DM particle mass. 

For the gamma-ray spectrum per annihilation ${\rm d}N_\gamma / {\rm d}E_\gamma$, we adopt two
characteristic spectra corresponding to annihilation with $100$\% branching ratios to 
$b\bar{b}$ and $\tau^+\tau^-$ final states. We use the {\tt PPPC4DMID} package 
\citep{Cirelli:2010xx} that is based on {\tt PYTHIA} (v8.135) and {\tt HERWIG} (v6.510) 
event generators. The spectra are dominated by emission from the decay of neutral pions. 
These are {\it primary} gamma-ray emissions, and should be distinguished from {\it secondary} 
emission that results from interactions of the annihilation products with the environment. 
Throughout this paper, we do not include secondary emissions. Secondary emissions
are only important for annihilation products propagating in regions of high baryon density, 
e.g., in and around galactic disks. This makes them interesting from the perspective of 
depending strongly on the astrophysical environment, but they are a minor contribution to 
the total gamma-ray yield from halos. 

\subsection{\label{subsec:halomodel}Cross-correlation with galaxy distribution}

Since the contribution from DM annihilation scales with the DM density squared,
any tracer of the DM density distribution should correlate with the gamma-ray
intensity distribution.
The galaxy distribution over the sky is among the most promising
tracers of the DM density distribution in the Universe.
While galaxies themselves are {\it biased} tracers of DM,
the statistical properties of galaxy clustering are well understood and
thus can be incorporated theoretically in the
so-called halo model approach, 
where one assumes that all the matter are contained in spherical DM halos.
Within the halo model approach, one can relate the galaxy distribution
to the DM density distribution
by introducing a halo occupation distribution (HOD).
The HOD $\langle N_{\rm gal}|M \rangle$ describes
the mean number of galaxies in a host halo of mass of $M$.
For a given $\langle N_{\rm gal}|M \rangle$ at redshift $z$, the
cross-power spectrum of galaxy overdensity $\delta_{\rm gal}$ and
DM overdensity squared $\delta^2$
can be expressed as \citep{Ando:2014aoa},
\beqa
P_{{\rm gal}, \delta^2}(k, z) 
&=& 
P^{1h}_{{\rm gal}, \delta^2}(k,z) + P^{2h}_{{\rm gal}, \delta^2}(k,z), 
\label{eq:3dpower}
\\
P^{1h}_{{\rm gal}, \delta^2}(k,z)
&=&
\int {\rm d}M\, \frac{{\rm d}n}{{\rm d}M}
\frac{{\cal J}(z,M)}{{\bar \rho}^2_{\rm dm}(z)}\frac{\langle N_{\rm gal}|M \rangle}{{\bar n}_{\rm gal}(z)}
\nonumber \\
&& 
\,\,\,\,\,\,\,\, \,\,\,\,
\times \, 
\tilde{u}_{\rm gal}(k|z, M) \tilde{u}_{\delta^2}(k|z, M), 
\label{eq:3dpower_1h} \\
P^{2h}_{{\rm gal}, \delta^2}(k,z)
&=&
\left[
\int {\rm d}M\, \frac{{\rm d}n}{{\rm d}M} b_{h}(z, M) 
\frac{\langle N_{\rm gal}|M \rangle}{{\bar n}_{\rm gal}(z)} 
\tilde{u}_{\rm gal}(k|z, M)
\right] \nonumber \\
&\times&
\left[
\int {\rm d}M\, \frac{{\rm d}n}{{\rm d}M} b_{h}(z, M) 
\frac{{\cal J}(z,M)}{{\bar \rho}^2_{\rm dm}(z)} 
\tilde{u}_{\delta^2}(k|z, M)
\right] \nonumber \\
&& 
\,\,\,\,\,\,\,\, \,\,\,\,
\times \, 
P_{\rm lin}(k, z),
\label{eq:3dpower_2h}
\eeqa
where 
$P_{\rm lin}(k,z)$ is the linear matter power spectrum at $z$,
${\rm d}n/{\rm d}M$ is the halo mass function, 
$b_{h}(z, M)$ represents the linear halo bias,
and 
${\cal J}(z, M)$ is the volume integral of DM density squared in the spherical halo.
We set the minimum halo mass to be $10^{-6}\, h^{-1}M_{\odot}$ in the integral in Eqs.~(\ref{eq:3dpower_1h}) and (\ref{eq:3dpower_2h}).
Note that the integral is insensitive to the minimum halo mass
as long as it is set to be $\simlt10^{10}\, h^{-1}M_{\odot}$ due to the functional form of the HOD for the LRGs.
In this paper, we define the term ${\cal J}\tilde{u}_{\delta^2}$ as
the fourier transform of $\rho^2_{\rm dm}({\bd x})$ inside a halo,
while $\langle N_{\rm gal}|M \rangle \tilde{u}_{\rm gal}$ corresponds to the Fourier transform 
of galaxy distribution in a host halo.
The mean number density of galaxies ${\bar n}_{\rm gal}(z)$ is then given by,
\beqa
{\bar n}_{\rm gal}(z) = \int {\rm d}M\, \frac{{\rm d}n}{{\rm d}M}\langle N_{\rm gal}|M \rangle.
\eeqa

To calculate $P_{{\rm gal}, \delta^2}$, 
we adopt the model of halo mass function and linear halo bias in Refs.~\cite{Tinker:2008ff, Tinker:2010my}.
We calculate $\tilde{u}_{\delta^2}$ following Ref.~\cite{Ando:2013ff}.
Apart from the HOD, the largest uncertainty in our halo model calculation is in the factor ${\cal J}$.
In particular, the amplitude of ${\cal J}$ is sensitive to the amount of substructures in a halo
and thus is still poorly known even for massive galaxies or cluster-size halos.
In practice, the ${\cal J}$ factor can be expressed as,
\beqa
{\cal J}(z, M) = (1+b_{sh}(z, M))\int {\rm d}V\, \rho^2_{h}(r|z, M),
\eeqa
where $\rho_{h}$ represents the density profile of a spherical halo 
and $b_{sh}$ is the boost factor, which describes 
the effective ``boost'' to the amplitude of the DM density squared
owing to subhalos.
For the smooth component of the density profile,
we adopt the NFW profile \cite{Navarro:1996gj}
with concentrations as given in Ref~\cite{Prada:2011jf}.
For the boost factor $b_{sh}$, we consider two extreme 
phenomenological models by Refs.~\cite{Gao:2011rf,Sanchez-Conde:2013yxa}.
In the first, subhalo properties are scaled as power-laws and extrapolated 
many orders of magnitudes to the smallest subhalos ($10^{-6} M_\odot$)
\cite{Gao:2011rf}. This procedure yields large boost factors, but it is rather 
uncertain whether such extrapolations are valid to such small halos. We
 thus adopt the model as an optimistic scenario. In the second, the halo 
concentration is extrapolated by a relation that flattens at small halo masses, 
yielding smaller boosts \cite{Sanchez-Conde:2013yxa}. We treat this as a
conservative scenario, since the concentration-mass relation derives from 
field halos rather than subhalos; for a given mass, subhalos are expected to 
be more concentrated than field halos, hence the boost factor should be 
higher. Ref.~\cite{Bartels:2015uba} estimates the effect is a factor of 2--5
increase in the boost factors. 

Observations of a large number of galaxies enable us to constrain the HOD
fairly precisely.
The HOD of luminous red galaxies has already been studied 
with number counts \cite{Reid:2008sy}, spatial clustering \cite{Wake:2008mf},
and galaxy-galaxy lensing analysis \cite{Hikage:2012zk}.
A popular HOD model is given by,
\beqa
\langle N_{\rm gal}|M \rangle 
&=& 
\langle N_{\rm cen}|M \rangle \left( 1+ \langle N_{\rm sat}|M \rangle \right), \\
\langle N_{\rm cen}|M \rangle 
&=& 
\frac{1}{2} \left[ 1+{\rm erf}\left(\frac{\log M-\log M_{\rm min}}{\sigma_{\log M}}\right)\right], \\
\langle N_{\rm sat}|M \rangle
&=&
\left(\frac{M-M_{\rm cut}}{M_{1}}\right)^{\alpha},
\eeqa
where $\langle N_{\rm cen}|M \rangle$ represents the HOD of central galaxies which resides
at the halo center
and $\langle N_{\rm sat}|M \rangle$ represents the contribution from satellite galaxies.
In this paper, we adopt the values of five parameters 
($M_{\rm min}$, $\sigma_{\log M}$, $M_{\rm cut}$, $M_{1}$, $\alpha$) in Ref.~\cite{Reid:2008sy}.
We also assume that the satellite galaxy distribution follows the DM distribution
within a halo.
Thus, $\langle N_{\rm gal}|M \rangle \tilde{u}_{\rm gal}$ is set to be,
\beqa
\langle N_{\rm gal}|M \rangle \tilde{u}_{\rm gal}
&=&
\langle N_{\rm cen}|M \rangle 
\nonumber \\
&& 
\,\,\,\,\,\,\,\, 
+ \,
\langle N_{\rm cen}|M \rangle \langle N_{\rm sat}|M \rangle \tilde{u}_{\rm dm},
\eeqa
where $\tilde{u}_{\rm dm}$ is the Fourier transform of $\rho_{h}/M$.

\begin{figure}[!t]
\begin{center}
       \includegraphics[clip, width=0.85\columnwidth]
       {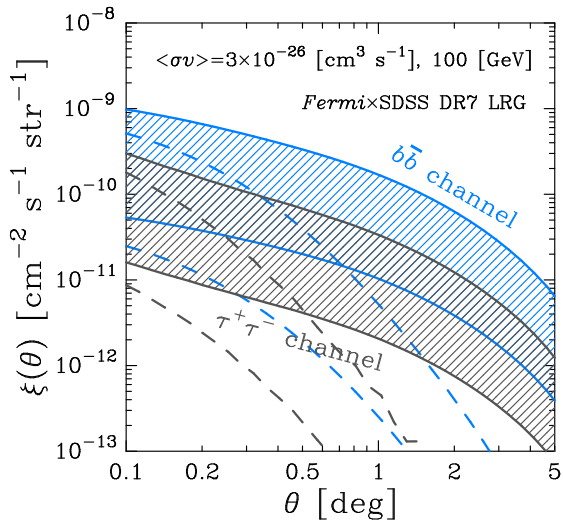}
     \caption{
     \label{fig:model_100GeV}
     The cross-correlation function of the EGB with LRGs.
     We plot the correlation expected from DM annihilation.
     We assume a DM particle mass of 100 GeV and canonical cross-section of
     $\langle \sigma v \rangle = 3\times 10^{-26}\, {\rm cm}^3/{\rm s}$.
     The different colors represent different annihilation channels: 
     $b\bar{b}$ (cyan) and $\tau^{+}\tau^{-}$ channel (gray).
     For each colored line, the shaded band indicates
     our conservative estimate of the theoretical uncertainty caused by 
     the subhalo boost factor model.
     The solid line shows the total correlation function
     while the dashed line represents the so-called one-halo term (see text).
  } 
    \end{center}
\end{figure}

Using the Limber approximation \cite{Limber:1954zz} and Eq.~(\ref{eq:Intensity}), 
we can calculate the angular cross-power spectrum of the LRG surface density 
and extragalactic gamma-rays emitted through DM annihilation as,
\beqa
C_{\rm gal, dm}(\ell) &=& 
\int \frac{{\rm d} \chi}{\chi^2} 
W_{\rm gal}(\chi) W_{\rm dm}(\chi) 
\nonumber \\
&&
\,\,\,\,\,\,\,\, \,\,\,\,\,\,\,\, 
\times \, 
P_{{\rm gal}, \delta^2}\left(k=\frac{\ell+1/2}{\chi}, z(\chi)\right), 
\eeqa
where $P_{{\rm gal}, \delta^2}$ is given by Eq.~(\ref{eq:3dpower})
and $W_{\rm gal}$ is defined as the normalized redshift distribution of galaxies \cite{Ando:2014aoa}.
Here, we define $W_{\rm dm}$ as,
\beqa
I_\gamma(\hat{\bd n}) &=& \int {\rm d}\chi W_{\rm dm}(\chi)\, \left(\frac{\rho_{\rm dm}(\chi\hat{\bd n}, \chi)}{\bar{\rho}_{\rm dm}}\right)^2,\\
W_{\rm dm}(\chi) &=& \frac{\langle \sigma v\rangle}{8\pi} 
\left(\frac{\bar{\rho}_{\rm dm}}{m_{\rm dm}}\right)^2
(1+z)^3 
\nonumber \\
&&
\,\,\,\,\,\,\,\, \,\,\,\,\,\,\,\, 
\times \,
\int_{E_{\gamma, {\rm min}}}^{E_{\gamma, {\rm max}}} {\rm d}E_{\gamma} \, 
\frac{{\rm d}N_{\gamma}}{{\rm d}E^{\prime}_{\gamma}}
\, e^{-\tau(E^{\prime}_{\gamma},z)},
\eeqa
where $E^{\prime}_{\gamma} = (1+z)E_{\gamma}$ 
[see Eq.~(1)]
and $I_\gamma(\hat{\bd n})$ is the gamma-ray intensity integrated over a given energy range of 
$E_{\gamma, {\rm min}}$ to $E_{\gamma, {\rm max}}$ along a direction $\hat{\bd n}$.
In practice, a more direct observable is the two-point cross-correlation function of 
the surface galaxy density $\Sigma_{\rm gal}(\hat{\bd n})$
and gamma-ray intensity $I_\gamma(\hat{\bd n})$.
The two-point cross-correlation function $\xi(\theta)$
can be calculated from the angular power spectrum by the following equations:
\beqa
\xi(\theta) 
&=& 
\langle I_{\gamma}(\hat{\bd n}) \Sigma_{\rm gal}(\hat{\bd n} + {\bd \theta})\rangle
-\langle I_{\gamma}\rangle \langle \Sigma_{\rm gal}\rangle
\nonumber \\
&=&
\sum_{\ell} \frac{2\ell + 1}{4\pi}C_{\rm gal, dm}(\ell)P_{\ell}(\cos \theta),
\eeqa
where $P_{\ell}(\cos \theta)$ is  the  Legendre polynomial.

Figure \ref{fig:model_100GeV} shows our benchmark model of $\xi(\theta)$.
We consider two representative annihilation channels ($b\bar{b}$ and $\tau^{+}\tau^{-}$)
for a 100 GeV DM with a thermal cross-section
$\langle \sigma v \rangle = 3\times 10^{-26}\, {\rm cm}^3/{\rm s}$.
We also show our conservative treatment of the theoretical uncertainty 
originating from the models of boost factor $b_{sh}$.
Hence, the expected correlation for each channel lies in the shaded region 
in Fig.~\ref{fig:model_100GeV}.
The solid line shows the sum of the one-halo term and the two-halo term,
whereas the dashed line represents the contribution of the one-halo term alone.
In our theoretical model, the uncertainty of the boost factor causes an uncertainty in $\xi(\theta)$
of a factor of $\sim10$.
Also, the one-halo term dominates on scale of $\simlt 0.2$ deg, while
the two-halo term induces significant correlations at $\simgt 1$ deg.
 
\section{\label{sec:data}DATA}

\subsection{\label{subsec:lrg}SDSS DR7 LRG}

The Slone Digital Sky Survey (SDSS) provides 
the largest sample of LRGs to date.
The SDSS has imaged the sky 
at high Galactic latitude in five passbands $u, g, r, i,$ and $z$ 
\cite{Fukugita:1996qt, Gunn:1998vh},
taken by the 2.5 m APO telescope \cite{Gunn:2006tw}. 
The process \cite{Lupton:2001zb, Stoughton:2002ae, Pier:2002iq, Ivezic:2004bf} 
and calibration \cite{Hogg:2001gc, Smith:2002pca, Tucker:2006dv} 
of observed images enable to select galaxies \cite{Eisenstein:2001cq, Strauss:2002dj}
and quasars \cite{Richards:2002bb} spectroscopically
with twin fiber-fed double spectrographs.
Targets  are  assigned  to  plug  plates 
according to the tiling algorithm in Ref~\cite{Blanton:2001yk}.
The SDSS I/II imaging surveys were 
completed with a seventh data release (DR7) \cite{Abazajian:2008wr}.

In Ref~\cite{Eisenstein:2001cq}, the authors developed an algorithm for
selecting LRGs that we use in the present paper as tracers of the underlying matter
in and around massive DM halos.
We use the publicly available ``Bright" LRG catalog made by Ref~\cite{Kazin:2009cj}.
The sample consists of 30,272 LRGs with absolute magnitude
$-23.2< M_g <-21.2$ in the redshift range of $0.16 < z <0.44$.
They are populated over about 7,200 square degrees in the Northern Galactic Cap
with nearly constant comoving number density.
The two-point correlation function of the LRG sample has been studied in Ref~\cite{Kazin:2009cj}.
We note that possible systematic effects on the correlation analysis is expected to be smaller than the sample variance estimated from mock LRG catalogues (see the appendix in Ref~\cite{Kazin:2009cj} for details).

Note that Ref~\cite{2015ApJS..217...15X} also considered the cross correlation of the EGB with LRGs.
The LRGs used in this paper are the spectroscopic LRG sample, 
while Ref~\cite{2015ApJS..217...15X} considered the photometric LRGs from SDSS data release 8.
We refer the former as spec-LRGs, 
while the latter is denoted as photo-LRGs.
The advantages to use spec-LRGs in our paper are as follows:
(i) the accurate redshift distribution by spectroscopic redshift,
(ii) the observationally constrained HOD,
and 
(iii) a large number of mock catalogs.
According to the accurate redshift distribution and 
the well-constrained HOD, 
the spec-LRGs are known for having relatively lower redshift than photo-LRGs and the typical halo mass of $\sim10^{14}\, h^{-1}M_{\odot}$.
The redshift range of the photo-LRGs in Ref~\cite{2015ApJS..217...15X} is $0.45<z<0.65$ with a mean redshift of 0.5, 
while the spec-LRGs locate at $0.16 < z <0.44$. 
Since more massive DM halos at lower redshift would have
larger contribution to the EGB through DM annihilation, 
we decide to use the spec-LRGs for cross-correlation analysis with EGB.
Furthermore, compared to Ref~\cite{2015ApJS..217...15X}, 
we improve the cross-correlation analysis 
by using accurate covariances measured directly
from a large set of mock catalogues of the spec-LRGs.

\subsection{\label{subsec:egb}Extragalactic gamma-ray background}

The \textit{Fermi} satellite provides a full-sky coverage of the GeV gamma-ray sky.
We utilize publicly available \textit{Fermi}-LAT Pass 7 Reprocessed photon data taken
from August 2008 to December 2014. We divide the SDSS LRGs survey patch into
two square ROIs, left and right, each covering $80^\circ$ on a side.
Using the Fermi Tools version {\tt v10r0p5}, we use the {\tt gtmktime}
tool to reduce the photon data by removing data taken during nonsurvey modes
and removing times when the satellite rocking angle exceeds $52^\circ$ with
respect to the zenith. These produce a photon sample suitable for analyses. 
We work with {\tt ULTRACLEAN}-class photons,
which are events that pass the most stringent quality cuts, and we use photons
between 1 and 500 GeV in energy. As described below, both choices help adopt
a small point source mask. With the {\tt gtbin} tool, we bin photons into
$0.2^\circ \times 0.2^\circ$ pixels and 30 energy bins equally spaced logarithmically.
These binnings are the recommended values by the Fermi collaboration to ensure
reasonable analysis outcomes. We generate exposure maps using the standard
{\tt gtltcube} and {\tt gtexpcube2} tools, using the {\tt P7REP\_ULTRACLEAN\_V15} instrument response function.

We obtain the extragalactic diffuse photons separately for each ROI. 
First, we perform a likelihood fit of the reduced photon counts cube
using the {\tt gtlike} tool and find the best fitting Galactic diffuse emission model. 
We include the following templates in the fits: all the known point and
diffuse sources in the 3FGL catalog, a template for the Galactic diffuse
foreground emission, and a template for the isotropic emission. 
As described in Section \ref{subsec:galprop}, we consider multiple templates
for the Galactic diffuse emission model, whose uncertainties can dominate
analyses of residuals. 
Only the normalizations of the Galactic diffuse and isotropic emission
templates are varied in the fits. 
We have confirmed that 
the best-fit Galactic diffuse maps between two ROIs 
change by $1\%$ or less and 
the difference is unimportant for our analysis.
Next, we create a model counts cube for the Galactic emission model,
as well as any known diffuse sources in the 3FGL catalog. 
These model counts cubes are subtracted from the raw photon counts cube.
Finally, we mask all known point sources within the ROI. 
Our default mask size is $1^\circ$ around each point source, and is
motivated by the behavior of the point spread function (PSF): 
the 68\% containment angle is
$\sim 0.9^\circ$ at 1 GeV and $\sim 0.26^\circ$ at 10 GeV when both
front and back conversion tracks are included. 
Since most point sources have steeply falling spectra, their emissions
are dominated by lower energy photons, and we conservatively chose a $1^\circ$ mask. 
We also test that our results are unaffected by changing the mask size to $2^\circ$.

For our cross-correlation analysis, 
we use both front and back conversion tracks.
To reduce the impact of the Galactic emission 
on our analysis, we apply a Galactic latitude cut $|b|>20^\circ$
\cite{Xia:2011ax}.   
Moreover,  we  also  exclude  the  region  associated to the
Fermi Bubbles and the Loop I structure
by applying a Galactic longitude cut $50^\circ < \ell < 280^\circ$
\cite{Su:2010qj}.  

\begin{figure*}
\begin{center}
       \includegraphics[clip, width=0.96\columnwidth]
       {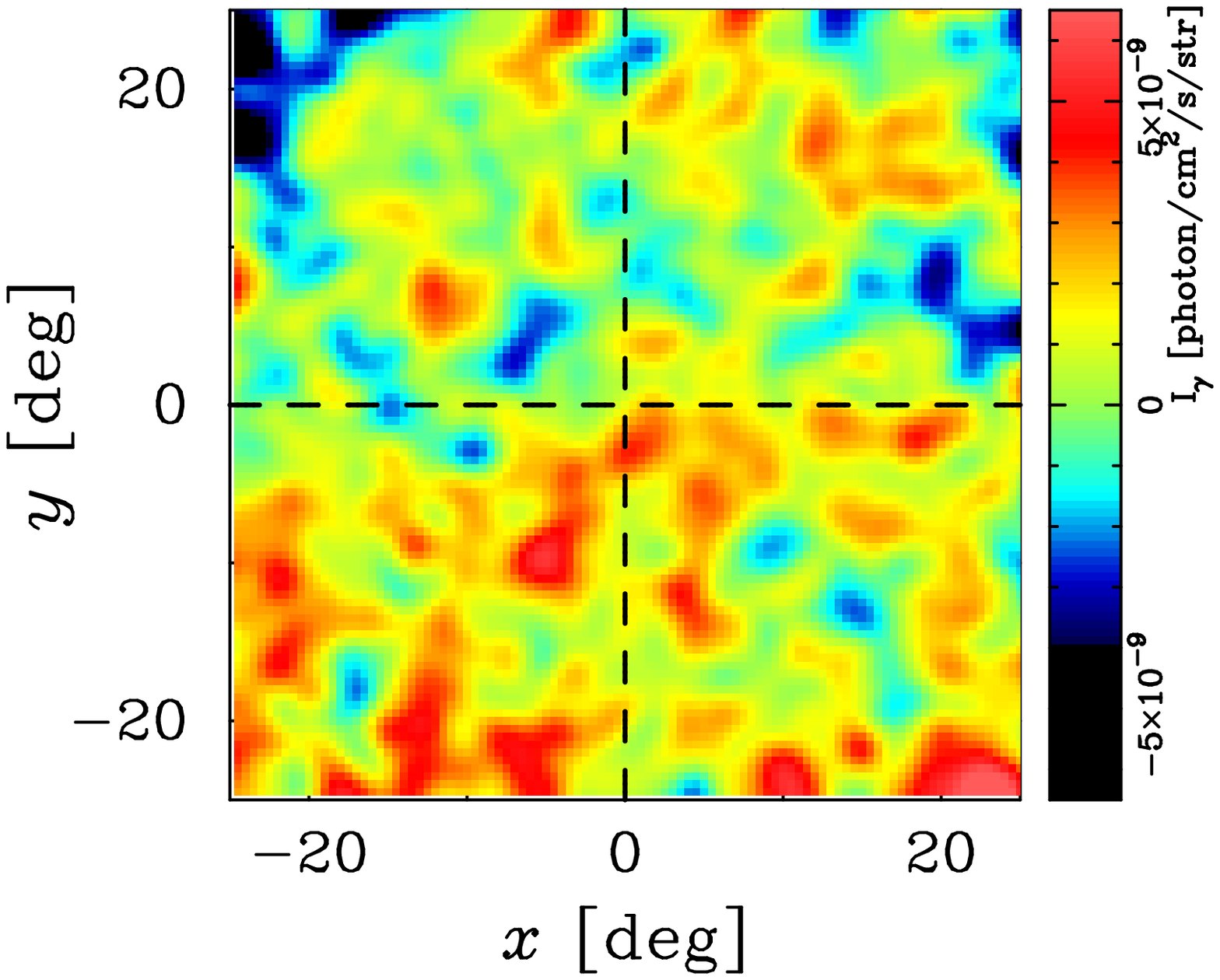}
       \includegraphics[clip, width=0.80\columnwidth]
       {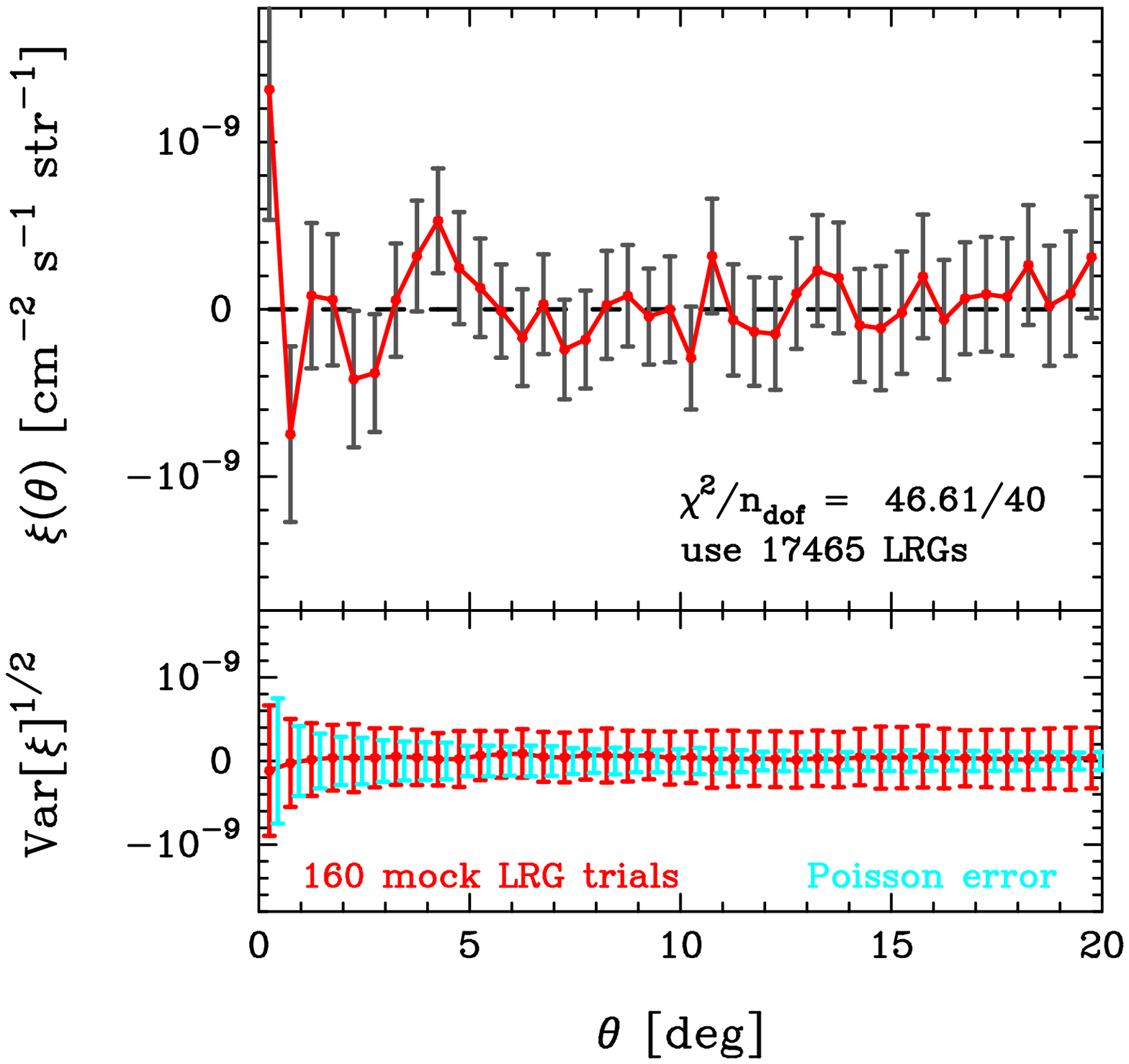}
     \caption{
     \label{fig:image_and_signal}
     The cross-correlation of the EGB with LRGs. {\it Left}:
     the stacked image of EGB photons around 17,465 LRGs.
     The color coordinate represents the EGB intensity in the energy range of 1--500 GeV.
     {\it Right}: the cross-correlation function as a function of separation angle.
     In the top panel, the red line represents our measurement and the gray error
     bars correspond to the statistical
     uncertainty estimated from 160 mock catalogs.
     The bottom panel shows the contribution to the statistical uncertainty from sample variance (red) and 
     Poisson photon noise (cyan) separately. } 
    \end{center}
\end{figure*}

\section{\label{sec:cross}CROSS-CORRELATION OF EGB WITH LRG}

In order to calculate the cross-correlation of the EGB with the LRGs, 
we use the following estimator: 
\beqa
\xi(\theta)
&=&
\left[
{\displaystyle \sum^{N_{\rm pix}}_{i}\sum^{N_{\rm pix}}_{j}
(I^{\rm obs}_\gamma({\bd \phi}_{i})-I^{\rm gm}_\gamma({\bd \phi}_{i}))
\Sigma_{\rm gal}({\bd \phi}_{j})\Delta_{\theta}({\bd \phi}_{i}-{\bd \phi}_{j})}
\right]
\nonumber \\
&&\times
\left[{\displaystyle \sum^{N_{\rm pixel}}_{i}\sum^{N_{\rm pixel}}_{j}\Delta_{\theta}({\bd \phi}_{i}-{\bd \phi}_{j})}\right]^{-1},
\label{eq:CCest}
\eeqa
where $N_{\rm pix}$ is the number of pixels in the gamma-ray intensity map,
$\Sigma_{\rm gal}({\bd \phi}_{i})$ is the overdensity of galaxies in pixel $i$,
$I^{\rm obs}_\gamma({\bd \phi}_{i})$ is the observed gamma-ray intensity in pixel $i$, and
$I^{\rm gm}_\gamma({\bd \phi}_{i})$ is the contribution from the Galactic emission model estimated 
using the ${\it Fermi}$-LAT diffuse template and detector modeling.
In Eq.~(\ref{eq:CCest}), we define the function $\Delta_{\theta}({\bd \phi}) = 1$ 
for $\theta-\Delta \theta/2 \le \phi \le \theta+\Delta \theta/2$ and zero otherwise.
We have checked that our estimator is consistent with a zero signal
when applied to mock LRG catalogues and the observed gamma-ray intensity map.

For binning in angular separation $\theta$, we use 40 bins linearly spaced
in $\Delta \theta = 0.5$ deg. In calculating Eq.~(\ref{eq:CCest}), 
we do not perform any corrections to the effect of PSF of the ${\it Fermi}$-LAT detector.
In the present paper,
we take the PSF smearing into account in theoretical models.

We then estimate the covariance matrix $C_{ij}$ of the estimator Eq.~(\ref{eq:CCest}) by,
\beqa
C_{ij} = \frac{1}{N_{\rm re}-1}\sum_{r}(\xi^{r}(\theta_{i})-{\bar \xi}(\theta_{i}))(\xi^{r}(\theta_{j})-{\bar \xi}(\theta_{j})),\label{eq:cov}
\eeqa
where $\xi^{r}(\theta_{i})$ 
is the estimator for the $i$-th angular bin obtained 
from the $r$-th realization, 
and $N_{\rm re} = 160$ is the number of randomized catalogues. 
The ensemble average of the $i$-th angular bin over 160 realizations, 
${\bar \xi}(\theta_{i})$, is 
simply given by,
\beqa
{\bar \xi}(\theta_{i})
=\frac{1}{N_{\rm re}}\sum_{r}\xi^{r}(\theta_{i}).
\eeqa
In this paper, we consider two sets of random catalogues:
mock LRG catalogues created from ``Las Damas" N-body simulations\footnote{
http://lss.phy.vanderbilt.edu/lasdamas/
} (McBride {\it et al}, in prep)
and randomized gamma-ray count maps generated by the real data through Poisson processes.
The former provides random realizations of spatial clusterings of LRGs,
while the latter can be used to simulate the photon count noise.
To simulate the photon count noise, 
we generate 160 randomized count maps in the same way as shown in Ref.~\cite{Shirasaki:2014noa}.
According to the estimated exposure over the SDSS region, 
we derive the gamma-ray intensity for each randomized photon map.
We then repeat the cross-correlation analysis for two different sets of catalogues,
i.e., 
160 randomized gamma-ray maps and the observed galaxy catalogue
or 
the observed gamma-ray map and the 160 mock galaxy catalogues.
We estimate the statistical error associated 
with the spatial clusterings of LRGs and 
the photon noise by summing these two contributions.
Note that our estimate of covariance includes the sampling variance of LRGs
because we use {\it independent} mock LRGs of which spatial clustering
pattern is closely matched to that of the real LRGs.
We calculate Eq.~(\ref{eq:cov}) 
on the assumption that the LRGs and the EGB
do not correlate each other. 
For our analysis, it is sufficient to
verify whether the LRGs correlate with the EGB or not. 

The cross-correlation estimator [Eq.~(\ref{eq:CCest})] is 
dependent on the model for the 
astrophysical foreground emission of our own Galaxy. 
We therefore use variants of the foreground emission models
provided by the ${\it Fermi}$ collaboration 
to assess the impacts on the estimated EGB.
We work with 35 different Galactic diffuse models from Ref.~\cite{Ackermann:2012pya}.
The details of these analyses are summarized in Section \ref{subsec:galprop}. 
  
\section{\label{sec:res}RESULT}

Figure \ref{fig:image_and_signal} summarizes the result of our cross-correlation analyses.
In the left panel, we show a stacked image of the EGB intensity around the angular positions of LRGs.
In the right panel,
we present the cross-correlation function as a function of separation angle $\theta$.
In the upper portion of the right panel, the red line shows the measured signal
and the gray error bars 
show the statistical error estimated from a set of mock catalogs.
In the lower portion of the right panel, we show the contributions to the
total statistical uncertainty from photon noise and sample variance separately,
and find that the photon noise is comparable to the sample variance of LRGs 
in our measurement.

In order to quantify the significance of the measured 
cross-correlation signal with respect to the statistical error, 
we use the $\chi^2$ statistics defined by,
\beqa
\chi^2 
= \sum_{i,j}\xi(\theta_{i})C^{-1}_{ij}\xi(\theta_{j}),
\eeqa
where ${\bd C}^{-1}$ denotes the inverse covariance matrix 
estimated from the randomized realization shown in Section~\ref{sec:cross}.
In our analysis, the number of degrees of freedom is 40.
The resulting value of $\chi^2 / n_{\rm dof}$ is 46.61/40,
which implies a good fit without DM annihilation.

We are now able to use the null detection of the cross-correlation to place 
constraints on the DM annihilation cross-section.
For this purpose, we use the maximum likelihood analysis.
We assume that the data vector ${\bd D}$ is well approximated 
by the multivariate Gaussian distribution with covariance ${\bd C}$.
In this case, $\chi^2$ statistics (log-likelihood) is given by,
\beqa
\chi^2({\bd p}) = \sum_{i,j}(D_{i}-\mu_{i}({\bd p}))C^{-1}_{ij}(D_{j}-\mu_{j}({\bd p})), \label{eq:logL}
\eeqa
where ${\bd \mu}({\bd p})$ is the theoretical template 
for the set of parameters of interest.
In this paper, we use the halo model approach described
in Section \ref{subsec:halomodel} to calculate the theoretical prediction.
As parameters of interest ${\bd p}$,  
we simply consider the DM particle mass and the annihilation 
cross-section, $m_{\rm dm}$ and $\langle \sigma v \rangle$.
The data vector ${\bd D}$ consists of the ten measured cross-correlation 
amplitudes in the range of $\theta=[0, 5]$ degree as,
\beqa
D_{i} = \{ \xi(\theta_{1}), \xi(\theta_{2}),..., \xi(\theta_{10}) \},
\eeqa
where $\theta_{i}$ is the $i$-th angular separation bin.
The inverse covariance matrix ${\bd C}^{-1}$ includes 
both the statistical error owing to the spatial clusterings of LRGs
and the photon Poisson error.
We consider the 95 \% confidence level of posterior distribution
function of parameters.
This is given by the contour line in the two dimensional space 
($m_{\rm dm}$ and $\langle \sigma v \rangle$), which is defined as 
\beqa
\Delta \chi^2({\bd p}) = \chi^2({\bd p})-\chi^2({\bd \mu}=0)=6.17.
\eeqa

As discussed in Section \ref{subsec:halomodel}, the choice of boost factor $b_{sh}$
affects the theoretical predictions significantly, by a factor of about ten.
We therefore derive constraints based on an optimistic 
scenario adopting Ref.~\cite{Gao:2011rf}, and a conservative 
scenario adopting Ref.~\cite{Sanchez-Conde:2013yxa}. 

\begin{figure}[!t]
\begin{center}
       \includegraphics[clip, width=0.90\columnwidth]
       {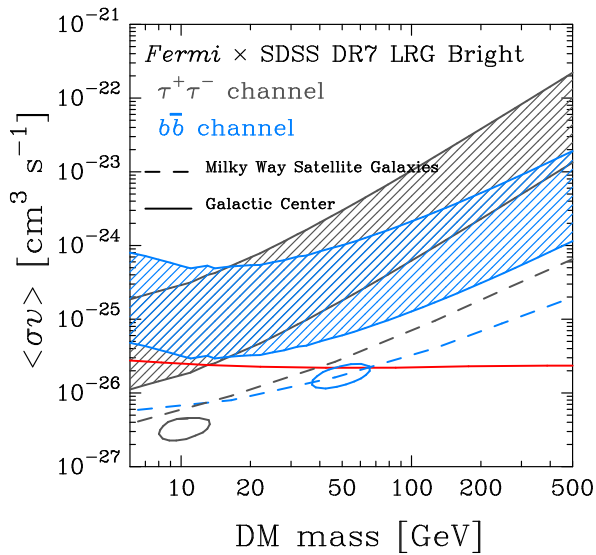}
     \caption{
     \label{fig:sigv_m_plane}
     Shown are 95\%C.L.~upper limits on the DM annihilation cross-section. The two shaded bands are the
     limits derived from our cross-correlation analysis of the EGB with LRGs, showing two annihilation channels, 
     $b\bar{b}$ (cyan) and $\tau^{+}\tau^{-}$ (gray).
     For each channel, the shaded band represents a conservative estimate of the theoretical uncertainty 
     due to DM substructure. 
     For comparison, we also plot constraints from Ref.~\cite{Ackermann:2013yva} derived from staked satellites 
     galaxies of the Milky Way (dashed), and regions of best fit from studies of the Milky Way Galactic center
     from Ref.~\cite{Abazajian:2014fta} (closed circles), respectively. 
     The red line shows the canonical cross-section expected for thermal relic DM \cite{2012PhRvD..86b3506S}.
  } 
    \end{center}
\end{figure}

Figure \ref{fig:sigv_m_plane} shows the result of our likelihood analysis on the
DM parameter space $m_{\rm dm}$ and $\langle \sigma v \rangle$.
We plot the constraints for two representative particle physics 
model, the $\tau^{+}\tau^{-}$ channel and the $b\bar{b}$ 
channel. The shaded band is derived from the two extreme assumptions of 
the boost factor, and thus the true bound is expected to lie in the
shaded region. The constraint for the large boost factor model \cite{Gao:2011rf} 
is significantly stronger, as expected. 
For reference, the red line indicates the canonical cross-section of
for a thermally produced DM \cite{2012PhRvD..86b3506S}. 

\begin{figure*}[!t]
\begin{center}
       \includegraphics[clip, width=0.90\columnwidth]
       {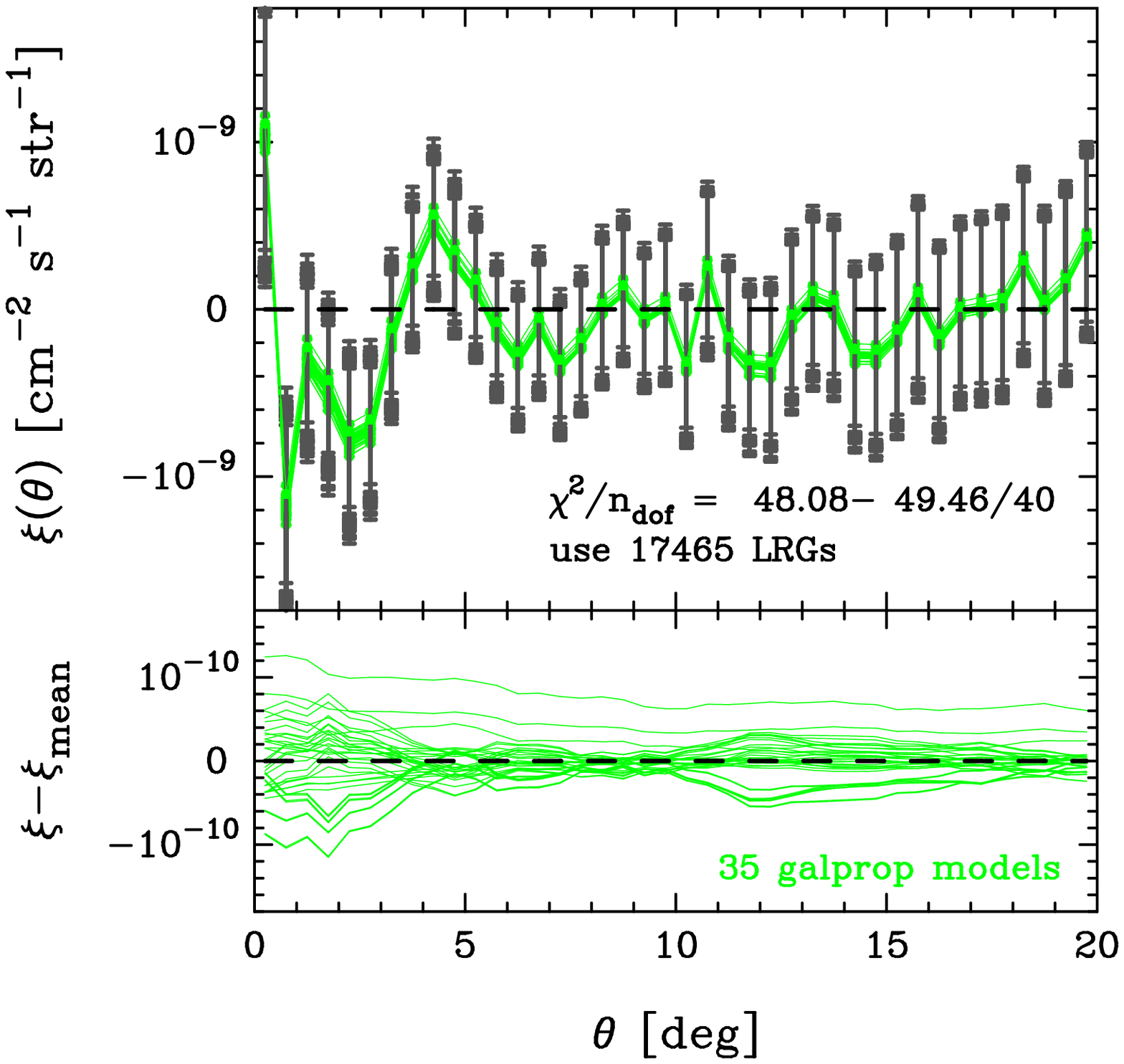}
       \includegraphics[clip, width=0.88\columnwidth]
       {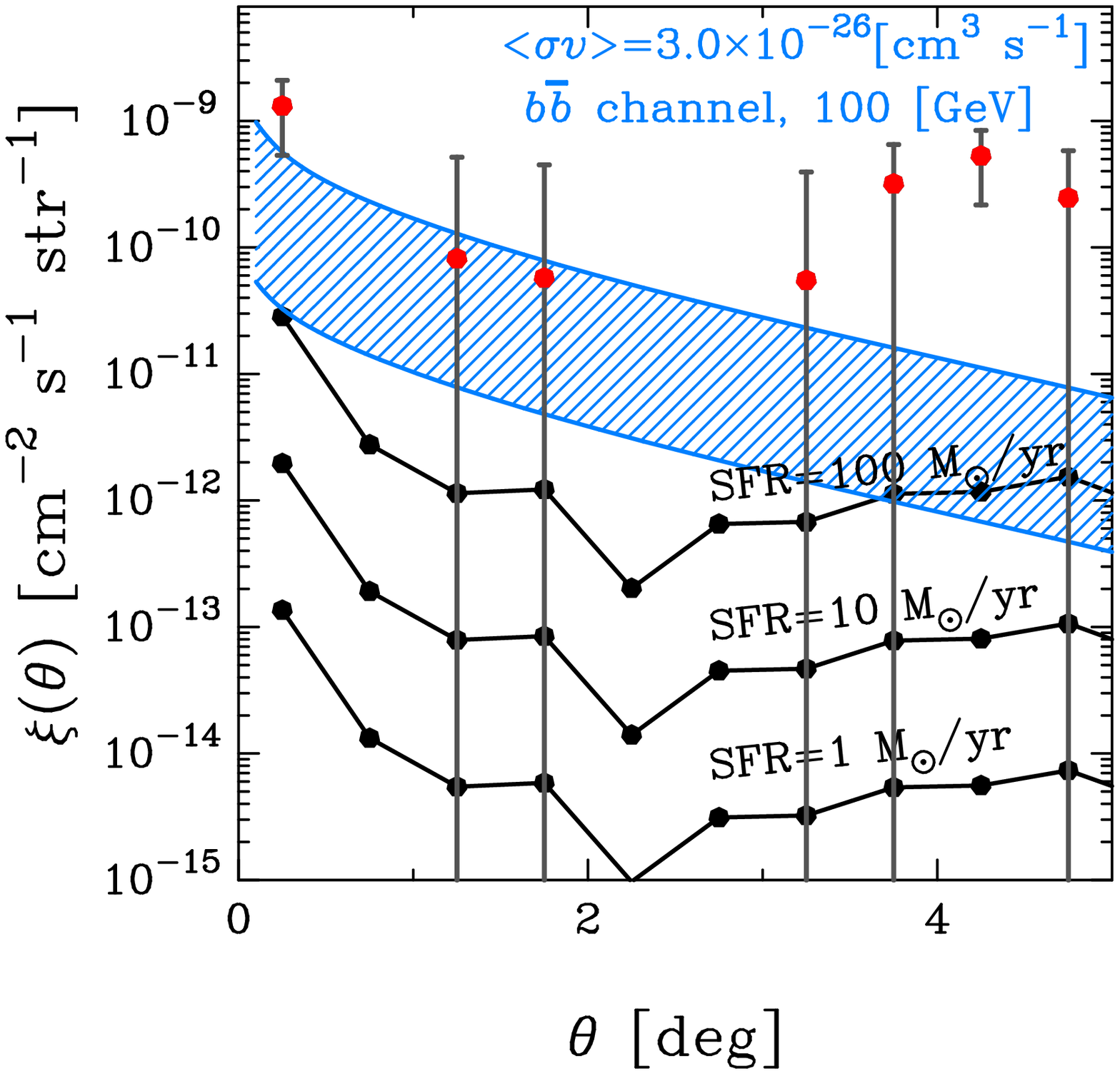}
     \caption{
     \label{fig:sys_err}
     Systematic uncertainties in our cross-correlation analysis.
     {\it Left}: the results of cross-correlation analyses for EGB data obtained
     using 35 different
     Galactic gamma-ray foreground templates. 
     In the top portion, the green lines and grey error bars show 
     the measured cross-correlation signals $\xi$
     for a given galactic template. The green lines
     are virtually indistinguishable.
     The bottom portion shows the differences in $\xi$ from $\xi_{\rm mean}$, 
     where $\xi_{\rm mean}$ represents the mean cross-correlation signal over the data set derived from 35
     different Galactic gamma-ray foreground templates.
     {\it Right}: the expected correlation from star-forming gamma-rays in LRGs.
     The three solid lines show the cases with star formation rates of 1, 10, and 100 $M_{\odot}/{\rm yr}$.
     For comparison, the shaded region represents the expected cross-correlation signals calculated from our halo model in the case of annihilating DM (into $b\bar{b}$) with the mass of 100 GeV  and with a thermal canonical cross-section.
     We also plot the measured signal (but only positive) 
     as shown in Fig.~\ref{fig:image_and_signal} 
     by the red point with gray error bar.
  } 
    \end{center}
\end{figure*}

\section{\label{sec:sys}SYSTEMATIC UNCERTAINTIES}

In this section, we investigate possible systematic uncertainties in our
cross-correlation analysis.
We consider two contributions: (i) uncertainty in modeling the Galactic gamma-rays foregrounds,
and (ii) possible correlation due to gamma-ray emission from
star formation in LRGs.

\subsection{\label{subsec:galprop}Galactic model template}

The main contribution to the $E_{\gamma}>100$ MeV gamma-ray sky arises from diffuse emissions 
produced by interactions of Galactic cosmic rays with the gas and radiation fields in the Galaxy.
Estimates of the EGB are crucially affected by this Galactic diffuse emission which 
need to be subtracted from the observed gamma-ray intensity.

Determining the Galactic diffuse emission requires treatments of multiple processes. 
Cosmic-ray interactions with the interstellar gas produce neutral
pions which subsequently decay to gamma-ray pairs. 
Interactions with gas also give rise to bremsstrahlung emission,
which can be a significant contribution in gas-rich environments of the Galaxy. 
Finally, cosmic-ray electrons up-scatter low-energy photons to 
gamma-ray energies via the inverse-Compton process.
At the same time, cosmic rays propagate diffusively or rectilinearly in the
magnetic field of the Milky Way. The final prediction therefore depends on
the source distribution of cosmic rays,
the injection spectra and composition of cosmic rays,
the distribution of interstellar gas, light, and magnetic field 
in the Milky Way, and the assumed propagation model. 

In order to estimate the possible uncertainties in such complex phenomena,
we follow Ref.~\cite{FermiLAT:2012aa} 
and generate multiple Galactic gamma-ray foreground models using the {\tt GALPROP}
(version 54) propagation code 
\cite{1999ICRC....4...52S, 1999ICRC....4..255S, 2000Ap&SS.272..247M}.
In Ref.~\cite{FermiLAT:2012aa}, a total of 128 models were explored by varying the 
following parameters:
the cosmic-ray source distribution,
the cosmic-ray confinement region, 
and assumptions affecting the interstellar gas distribution.
For the cosmic-ray source distribution, four distributions were explored:
the supernova remnant distribution of Ref.~\cite{Case:1998qg},
the O-star distribution of Ref.~\cite{Bronfman:2000tw}, and the pulsar distributions 
of Refs.~\cite{Lorimer:2006qs} and \cite{Yusifov:2004fr}. For the cosmic-ray
confinement region, cylindrical volumes with heights $z=4$, 6, 8, and 10 kpc, 
and radii $R=20$ and 30 kpc were explored. For the interstellar gas distribution, 
two assumptions for the optical depth correction to the atomic hydrogen component
($T_s = 150$ K and $10^5$ K),
and two values for the cut at which dust emission is no longer used to 
correct for missing neutral gas ($E(B-V) = 2$ and 5 mags) were explored. 
For each of these models, parameters affecting the cosmic ray
and gas are varied and fitted to \textit{Fermi}-LAT data. The fitted parameters include
those setting the injected cosmic-ray 
nuclei and electron spectral shapes and normalizations,
the diffusion coefficient, as well as conversion factors between CO and molecular
hydrogen. 
We refer the reader to Ref.~\cite{FermiLAT:2012aa} for details of the parameters
and fits. 

We simulate a total of 35 among these 128 models. Specifically, 
we simulate all 32 models that use the 
pulsar distribution of Ref.~\cite{Lorimer:2006qs} as a proxy for the cosmic-ray
spatial distribution. In addition, we
simulate one additional model for each of the other three cosmic-ray source
distributions, all for the same inputs parameters $z=6$ kpc, $R=30$ kpc, $T_s=150$ K, 
and $E(B-V) = 2$ mag cut. 
We use these 35 model templates and repeat the cross-correlation 
analysis described in Section~\ref{sec:cross}.
We estimate not only the cross-correlation signal but also the statistical 
error for each galactic model template.
We therefore obtain 35 different binned signals ($\xi$) as functions 
of angular separation.

The left panel of Fig.~\ref{fig:sys_err} shows the results of the cross-correlation
analyses using the model templates. In the top panel, we plot {\it all} the cross-correlation 
signals with their respective statistical errors. 
The green lines appear virtually indistinguishable. 
Clearly, the differences among model templates are significantly smaller than the
statistical error.
The resulting $\chi^2$ ranges from 48.08 to 49.46 with 40 data bins.
The bottom panel shows the variance of $\xi$ among the 35 trials.
The typical amplitude of variance due to the uncertainty in the Galactic model templates is 
$\sim1\times10^{-10} \, {\rm cm}^{-2}\, {\rm s}^{-1}\, {\rm str}^{-1}$,
which is $\sim10$ times smaller than the current statistical uncertainty 
(see the right panel in Figure~\ref{fig:image_and_signal}).
Therefore, we conclude that the modeling of the Galactic gamma-ray distribution does not
significantly affect our result. This is a result of the SDSS survey region being at relatively
high latitudes where the Galactic diffusion emission models are relatively better modeled
than, e.g., along the Galactic plane \cite{FermiLAT:2012aa}. 

\subsection{\label{subsec:astrocon}Contribution from astrophysical sources}

There is another important uncertainty in studying DM annihilation with 
cross-correlation: gamma-rays emitted from star-forming or/and radio galaxies 
contaminating the cross-correlation measurement.
Generally, astrophysical sources are expected to contribute considerably to the EGB
\cite{Inoue:2011bm, Ackermann:2012vca, 2014ApJ...780..161D, 2015ApJ...800L..27A}.
In order to estimate the possible correlation of LRGs and gamma rays from astrophysical sources,
we utilize the observed LRG distribution and known scaling relations 
between the gamma-ray intensity and star-forming activity.

We consider several fixed values of the star-forming rate (SFR) of the LRGs, and
assign gamma-ray luminosity to each LRG
according to the gamma ray to SFR
scaling relation given in Ref~\cite{Ackermann:2012vca}.
The model yields correlations of the LRGs and astrophysical gamma rays
that are equal to the auto correlation of the LRGs multiplied by the gamma-ray intensity
of each astrophysical source.
Hence, the contribution from astrophysical sources $\xi_{\rm ast}(\theta)$ is,
\beqa
\xi_{\rm ast}({\rm SFR}, \theta) \simeq I_{\gamma, {\rm ast}}({\rm SFR}) \xi_{\rm LRG}(\theta),
\label{eq:xi_ast}
\eeqa
where $I_{\gamma, {\rm ast}}({\rm SFR})$ represents the gamma-ray intensity as a function of SFR
expected from the scaling relation 
and $\xi_{\rm LRG}(\theta)$ is angular correlation function of LRGs.
The right panel in Fig.~\ref{fig:sys_err} shows three different cases of $\xi_{\rm ast}(\theta)$ 
with SFR of 1, 10, and 100 $M_{\odot}/{\rm yr}$.
The SFR distribution of LRG systems has been studied in Ref~\cite{2011MNRAS.417..453D}
and the typical value ranges from 0.1 to 1 $M_{\odot}/{\rm yr}$.
Since we do not take into account the off-centering effect of astrophysical sources in the
LRG host halo or
the smearing effect of gamma-ray PSF, our estimation of $\xi_{\rm ast}$ with Eq.~(\ref{eq:xi_ast})
is conservative.
Clearly, the contribution from star-forming galaxies is small,
and thus does not affect our cross-correlation analysis of the EGB with LRGs.

Similar results are obtained in the case of radio galaxies.
Ref.~\cite{2009AJ....138..900H} presents a possible correlation of radio galaxies and LRGs,
and also shows that the typical radio luminosity of LRGs at 1.4 GHz 
is $L_{\rm 1.4 GHz}\sim4\times10^{22}\, {\rm W}\, {\rm Hz}^{-1}$.
Using the scaling relation in Ref.~\cite{Ackermann:2012vca}, this corresponds to an
expected gamma-ray 
intensity of $\sim3\times 10^{40}\, {\rm erg}\, {\rm s}^{-1}$. Scaling to SFR,
this is equivalent to
$\sim$20 $M_{\odot}/{\rm yr}$.
Hence, the contribution from radio galaxies is also not significant.

\section{\label{sec:con}CONCLUSION AND DISCUSSION}

In this paper, we have performed a cross-correlation analysis of the EGB with LRGs. 
Using gamma-ray data from the ${\it Fermi}$ satellite and 
17,465 LRGs in the SDSS catalog, we find that the cross-correlation
signal is consistent with null detection (Fig.~\ref{fig:image_and_signal}). 
By using up-to-date theoretical models based on recent galaxy clustering
and weak lensing studies, 
we estimate the statistical errors from a combination of real data
and a large set of mock
observations. We derive constraints on the DM annihilation cross-section, 
considering different DM
annihilation channels and different models of the amount of substructures in DM halos. 
The DM annihilation cross-section must be smaller than  
$\langle \sigma v \rangle < 10^{-25}-10^{-23} \ {\rm cm}^{3} \ {\rm s}^{-1}$ 
for a $100$ GeV DM, depending on the assumed parameters and annihilation channel
(Fig.~\ref{fig:sigv_m_plane}). 
The constraint improves for smaller DM mass. 

We have further investigated two systematic uncertainties in our DM constraint (Figure \ref{fig:sys_err}).
The first is due to the uncertainty in the Galactic gamma-ray foreground emission model.
In order to evaluate the overall error, we utilize 35 different Galactic diffuse emission templates 
produced by the {\tt GALPROP} code and 
repeat the cross-correlation analysis 
for the EGB data set derived by each template.
The variation in the correlation signal due to differences in the Galactic templates
is about ten times smaller than the statistical error of current observations.
Although the template uncertainties depend on sky position, the systematic 
uncertainty is thus expected to be unimportant even in future galaxy surveys with 
sky coverages of 20,000 square degrees that would reduce the statistical
error by a factor of $\sim\sqrt{20000/7200}=1.7$.
The second source of systematic error is the possible correlation due to gamma rays emitted from 
astrophysical sources associated with LRGs.
We estimate this contamination by using 
empirical scaling relations observed between the gamma-ray luminosity and star-forming activities of nearby 
galaxies \cite{Ackermann:2012vca}.
The estimated correlation is $20-30$ times smaller than 
the expected signal of DM annihilation with DM mass of 100 GeV.
Therefore, we conclude that LRGs are an ideal target for the {\it statistical} detection of DM annihilation
by means of cross-correlation analysis. 

Encouraged by the results of our systematic uncertainty investigations, 
we forecast the improvement expected with 
upcoming galaxy surveys, such as the LSST with a sky coverage of 20,000 square degrees. 
When we simply assume that the statistical uncertainty is reduced by a factor of $\sqrt{20000/7200}$,
the expected constraints on $\langle \sigma v \rangle$ 
would reach $7.8-130\times10^{-26} \ {\rm cm}^{3} \ {\rm s}^{-1}$ for the $b \bar{b}$ 
channel and $3.7-62\times10^{-25} \ {\rm cm}^{3} \ {\rm s}^{-1}$ for the $\tau^{+}\tau^{-}$ 
channel, both for a 100 GeV DM. 
For a lighter DM motivated by the Galactic center excess (e.g., \cite{Goodenough:2009gk,Hooper:2010mq,Hooper:2011ti,Boyarsky:2010dr,Abazajian:2012pn,Gordon:2013vta,Abazajian:2014fta,Daylan:2014rsa,Abazajian:2014hsa}),
the constraints would reach 
$2.84-42.1\times10^{-26} \ {\rm cm}^{3} \ {\rm s}^{-1}$ for the $b \bar{b}$ 
channel (assuming $40$ GeV mass) 
and $1.02-17.17\times10^{-26} \ {\rm cm}^{3} \ {\rm s}^{-1}$ 
for the $\tau^{+}\tau^{-}$ (assuming $10$ GeV mass).
Therefore, future large-area galaxy surveys will test the DM origin hypothesis of the Galactic center excess.
Note that the expected constraints by LRGs in 20,000 square degrees 
would be competitive to the recent constraints derived from the cross-correlation analysis with local galaxies \cite{2015PhRvL.114x1301R},
and it would be $\sim10$ times tighter than the current limits obtained 
from the energy spectrum of EGB (the conservative limit shown in Ref~\cite{Ackermann:2015tah}).

We have shown that the LRGs are a promising target to search for DM annihilation.
The single largest uncertainty is the amount of substructures in DM halos, which 
significantly affects our theoretical model of gamma-ray emission.
Previous works have cultivated theoretical understanding 
of the effect of halo substructures on the DM annihilation signal 
\cite{2008ApJ...686..262K, 2010PhRvD..81d3532K,2014MNRAS.441.1329Z,Bartels:2015uba},
but there has been limited information about the amount of substructures in cluster-sized DM halos.
Further observations will help understand the properties and
abundance of halo substructure to calibrate theoretical models.
Gravitational lensing analysis in nearby clusters \cite{2010ApJ...713..291O}, 
and stacking analysis of member galaxies in clusters \cite{2014MNRAS.438.2864L, 2015ApJ...799..188S},
would be the first base of such studies.

\begin{acknowledgements}
The authors thank 
Shin'ichiro Ando, Stefano Camera and Matteo Viel
for useful discussions and comments on the manuscript.
M.S. is supported by Research Fellowships of the Japan Society for 
the Promotion of Science (JSPS) for Young Scientists.
N.Y. acknowledges financial support from JST CREST.
Numerical computations presented in this paper were in part carried out
on the general-purpose PC farm at Center for Computational Astrophysics,
CfCA, of National Astronomical Observatory of Japan.
\end{acknowledgements}

\bibliography{ref_prd}

\begin{thebibliography}{79}%
\makeatletter
\providecommand \@ifxundefined [1]{%
 \@ifx{#1\undefined}
}%
\providecommand \@ifnum [1]{%
 \ifnum #1\expandafter \@firstoftwo
 \else \expandafter \@secondoftwo
 \fi
}%
\providecommand \@ifx [1]{%
 \ifx #1\expandafter \@firstoftwo
 \else \expandafter \@secondoftwo
 \fi
}%
\providecommand \natexlab [1]{#1}%
\providecommand \enquote  [1]{``#1''}%
\providecommand \bibnamefont  [1]{#1}%
\providecommand \bibfnamefont [1]{#1}%
\providecommand \citenamefont [1]{#1}%
\providecommand \href@noop [0]{\@secondoftwo}%
\providecommand \href [0]{\begingroup \@sanitize@url \@href}%
\providecommand \@href[1]{\@@startlink{#1}\@@href}%
\providecommand \@@href[1]{\endgroup#1\@@endlink}%
\providecommand \@sanitize@url [0]{\catcode `\\12\catcode `\$12\catcode
  `\&12\catcode `\#12\catcode `\^12\catcode `\_12\catcode `\%12\relax}%
\providecommand \@@startlink[1]{}%
\providecommand \@@endlink[0]{}%
\providecommand \url  [0]{\begingroup\@sanitize@url \@url }%
\providecommand \@url [1]{\endgroup\@href {#1}{\urlprefix }}%
\providecommand \urlprefix  [0]{URL }%
\providecommand \Eprint [0]{\href }%
\providecommand \doibase [0]{http://dx.doi.org/}%
\providecommand \selectlanguage [0]{\@gobble}%
\providecommand \bibinfo  [0]{\@secondoftwo}%
\providecommand \bibfield  [0]{\@secondoftwo}%
\providecommand \translation [1]{[#1]}%
\providecommand \BibitemOpen [0]{}%
\providecommand \bibitemStop [0]{}%
\providecommand \bibitemNoStop [0]{.\EOS\space}%
\providecommand \EOS [0]{\spacefactor3000\relax}%
\providecommand \BibitemShut  [1]{\csname bibitem#1\endcsname}%
\let\auto@bib@innerbib\@empty
\bibitem [{\citenamefont {{Jungman}}\ \emph {et~al.}(1996)\citenamefont
  {{Jungman}}, \citenamefont {{Kamionkowski}},\ and\ \citenamefont
  {{Griest}}}]{1996PhR...267..195J}%
  \BibitemOpen
  \bibfield  {author} {\bibinfo {author} {\bibfnamefont {G.}~\bibnamefont
  {{Jungman}}}, \bibinfo {author} {\bibfnamefont {M.}~\bibnamefont
  {{Kamionkowski}}}, \ and\ \bibinfo {author} {\bibfnamefont {K.}~\bibnamefont
  {{Griest}}},\ }\href {\doibase 10.1016/0370-1573(95)00058-5} {\bibfield
  {journal} {\bibinfo  {journal} {Phys.Rept.}\ }\textbf {\bibinfo {volume}
  {267}},\ \bibinfo {pages} {195} (\bibinfo {year} {1996})},\ \Eprint
  {http://arxiv.org/abs/hep-ph/9506380} {hep-ph/9506380} \BibitemShut {NoStop}%
\bibitem [{\citenamefont {Ackermann}\ \emph {et~al.}(2014)\citenamefont
  {Ackermann} \emph {et~al.}}]{Ackermann:2013yva}%
  \BibitemOpen
  \bibfield  {author} {\bibinfo {author} {\bibfnamefont {M.}~\bibnamefont
  {Ackermann}} \emph {et~al.} (\bibinfo {collaboration} {Fermi-LAT
  Collaboration}),\ }\href {\doibase 10.1103/PhysRevD.89.042001} {\bibfield
  {journal} {\bibinfo  {journal} {Phys.Rev.}\ }\textbf {\bibinfo {volume}
  {D89}},\ \bibinfo {pages} {042001} (\bibinfo {year} {2014})},\ \Eprint
  {http://arxiv.org/abs/1310.0828} {arXiv:1310.0828 [astro-ph.HE]} \BibitemShut
  {NoStop}%
\bibitem [{\citenamefont {{Buckley}}\ \emph {et~al.}(2015)\citenamefont
  {{Buckley}}, \citenamefont {{Charles}}, \citenamefont {{Gaskins}},
  \citenamefont {{Brooks}}, \citenamefont {{Drlica-Wagner}}, \citenamefont
  {{Martin}},\ and\ \citenamefont {{Zhao}}}]{2015PhRvD..91j2001B}%
  \BibitemOpen
  \bibfield  {author} {\bibinfo {author} {\bibfnamefont {M.~R.}\ \bibnamefont
  {{Buckley}}}, \bibinfo {author} {\bibfnamefont {E.}~\bibnamefont
  {{Charles}}}, \bibinfo {author} {\bibfnamefont {J.~M.}\ \bibnamefont
  {{Gaskins}}}, \bibinfo {author} {\bibfnamefont {A.~M.}\ \bibnamefont
  {{Brooks}}}, \bibinfo {author} {\bibfnamefont {A.}~\bibnamefont
  {{Drlica-Wagner}}}, \bibinfo {author} {\bibfnamefont {P.}~\bibnamefont
  {{Martin}}}, \ and\ \bibinfo {author} {\bibfnamefont {G.}~\bibnamefont
  {{Zhao}}},\ }\href {\doibase 10.1103/PhysRevD.91.102001} {\bibfield
  {journal} {\bibinfo  {journal} {Phys.Rev.}\ ,\ \bibinfo {eid} {102001}}
  (\bibinfo {year} {2015})},\ \Eprint {http://arxiv.org/abs/1502.01020}
  {arXiv:1502.01020 [astro-ph.HE]} \BibitemShut {NoStop}%
\bibitem [{\citenamefont {Ackermann}\ \emph
  {et~al.}(2015{\natexlab{a}})\citenamefont {Ackermann} \emph
  {et~al.}}]{Ackermann:2015zua}%
  \BibitemOpen
  \bibfield  {author} {\bibinfo {author} {\bibfnamefont {M.}~\bibnamefont
  {Ackermann}} \emph {et~al.} (\bibinfo {collaboration} {Fermi-LAT}),\
  }\href@noop {} {\  (\bibinfo {year} {2015}{\natexlab{a}})},\ \Eprint
  {http://arxiv.org/abs/1503.02641} {arXiv:1503.02641 [astro-ph.HE]}
  \BibitemShut {NoStop}%
\bibitem [{\citenamefont {Abazajian}\ and\ \citenamefont
  {Kaplinghat}(2012)}]{Abazajian:2012pn}%
  \BibitemOpen
  \bibfield  {author} {\bibinfo {author} {\bibfnamefont {K.~N.}\ \bibnamefont
  {Abazajian}}\ and\ \bibinfo {author} {\bibfnamefont {M.}~\bibnamefont
  {Kaplinghat}},\ }\href {\doibase 10.1103/PhysRevD.86.083511} {\bibfield
  {journal} {\bibinfo  {journal} {Phys.Rev.}\ }\textbf {\bibinfo {volume}
  {D86}},\ \bibinfo {pages} {083511} (\bibinfo {year} {2012})},\ \Eprint
  {http://arxiv.org/abs/1207.6047} {arXiv:1207.6047 [astro-ph.HE]} \BibitemShut
  {NoStop}%
\bibitem [{\citenamefont {Abazajian}\ \emph {et~al.}(2014)\citenamefont
  {Abazajian}, \citenamefont {Canac}, \citenamefont {Horiuchi},\ and\
  \citenamefont {Kaplinghat}}]{Abazajian:2014fta}%
  \BibitemOpen
  \bibfield  {author} {\bibinfo {author} {\bibfnamefont {K.~N.}\ \bibnamefont
  {Abazajian}}, \bibinfo {author} {\bibfnamefont {N.}~\bibnamefont {Canac}},
  \bibinfo {author} {\bibfnamefont {S.}~\bibnamefont {Horiuchi}}, \ and\
  \bibinfo {author} {\bibfnamefont {M.}~\bibnamefont {Kaplinghat}},\ }\href
  {\doibase 10.1103/PhysRevD.90.023526} {\bibfield  {journal} {\bibinfo
  {journal} {Phys. Rev.}\ }\textbf {\bibinfo {volume} {D90}},\ \bibinfo {pages}
  {023526} (\bibinfo {year} {2014})},\ \Eprint {http://arxiv.org/abs/1402.4090}
  {arXiv:1402.4090 [astro-ph.HE]} \BibitemShut {NoStop}%
\bibitem [{\citenamefont {Camera}\ \emph {et~al.}(2013)\citenamefont {Camera},
  \citenamefont {Fornasa}, \citenamefont {Fornengo},\ and\ \citenamefont
  {Regis}}]{Camera:2012cj}%
  \BibitemOpen
  \bibfield  {author} {\bibinfo {author} {\bibfnamefont {S.}~\bibnamefont
  {Camera}}, \bibinfo {author} {\bibfnamefont {M.}~\bibnamefont {Fornasa}},
  \bibinfo {author} {\bibfnamefont {N.}~\bibnamefont {Fornengo}}, \ and\
  \bibinfo {author} {\bibfnamefont {M.}~\bibnamefont {Regis}},\ }\href
  {\doibase 10.1088/2041-8205/771/1/L5} {\bibfield  {journal} {\bibinfo
  {journal} {Astrophys.J.}\ }\textbf {\bibinfo {volume} {771}},\ \bibinfo
  {pages} {L5} (\bibinfo {year} {2013})},\ \Eprint
  {http://arxiv.org/abs/1212.5018} {arXiv:1212.5018 [astro-ph.CO]} \BibitemShut
  {NoStop}%
\bibitem [{\citenamefont {{Fornengo}}\ and\ \citenamefont
  {{Regis}}(2014)}]{2014FrP.....2....6F}%
  \BibitemOpen
  \bibfield  {author} {\bibinfo {author} {\bibfnamefont {N.}~\bibnamefont
  {{Fornengo}}}\ and\ \bibinfo {author} {\bibfnamefont {M.}~\bibnamefont
  {{Regis}}},\ }\href {\doibase 10.3389/fphy.2014.00006} {\bibfield  {journal}
  {\bibinfo  {journal} {Frontiers in Physics}\ }\textbf {\bibinfo {volume}
  {2}},\ \bibinfo {eid} {6} (\bibinfo {year} {2014})},\ \Eprint
  {http://arxiv.org/abs/1312.4835} {arXiv:1312.4835 [astro-ph.CO]} \BibitemShut
  {NoStop}%
\bibitem [{\citenamefont {{Ando}}\ \emph {et~al.}(2014)\citenamefont {{Ando}},
  \citenamefont {{Benoit-L{\'e}vy}},\ and\ \citenamefont
  {{Komatsu}}}]{2014PhRvD..90b3514A}%
  \BibitemOpen
  \bibfield  {author} {\bibinfo {author} {\bibfnamefont {S.}~\bibnamefont
  {{Ando}}}, \bibinfo {author} {\bibfnamefont {A.}~\bibnamefont
  {{Benoit-L{\'e}vy}}}, \ and\ \bibinfo {author} {\bibfnamefont
  {E.}~\bibnamefont {{Komatsu}}},\ }\href {\doibase 10.1103/PhysRevD.90.023514}
  {\bibfield  {journal} {\bibinfo  {journal} {Phys.Rev.}\ ,\ \bibinfo {eid}
  {023514}} (\bibinfo {year} {2014})},\ \Eprint
  {http://arxiv.org/abs/1312.4403} {arXiv:1312.4403} \BibitemShut {NoStop}%
\bibitem [{\citenamefont {Ando}(2014)}]{Ando:2014aoa}%
  \BibitemOpen
  \bibfield  {author} {\bibinfo {author} {\bibfnamefont {S.}~\bibnamefont
  {Ando}},\ }\href {\doibase 10.1088/1475-7516/2014/10/061} {\bibfield
  {journal} {\bibinfo  {journal} {JCAP}\ }\textbf {\bibinfo {volume} {1410}},\
  \bibinfo {pages} {061} (\bibinfo {year} {2014})},\ \Eprint
  {http://arxiv.org/abs/1407.8502} {arXiv:1407.8502 [astro-ph.CO]} \BibitemShut
  {NoStop}%
\bibitem [{\citenamefont {{Camera}}\ \emph {et~al.}(2015)\citenamefont
  {{Camera}}, \citenamefont {{Fornasa}}, \citenamefont {{Fornengo}},\ and\
  \citenamefont {{Regis}}}]{2015JCAP...06..029C}%
  \BibitemOpen
  \bibfield  {author} {\bibinfo {author} {\bibfnamefont {S.}~\bibnamefont
  {{Camera}}}, \bibinfo {author} {\bibfnamefont {M.}~\bibnamefont {{Fornasa}}},
  \bibinfo {author} {\bibfnamefont {N.}~\bibnamefont {{Fornengo}}}, \ and\
  \bibinfo {author} {\bibfnamefont {M.}~\bibnamefont {{Regis}}},\ }\href
  {\doibase 10.1088/1475-7516/2015/06/029} {\bibfield  {journal} {\bibinfo
  {journal} {JCAP}\ }\textbf {\bibinfo {volume} {6}},\ \bibinfo {eid} {029}
  (\bibinfo {year} {2015})},\ \Eprint {http://arxiv.org/abs/1411.4651}
  {arXiv:1411.4651} \BibitemShut {NoStop}%
\bibitem [{\citenamefont {{Ajello}}\ \emph {et~al.}(2015)\citenamefont
  {{Ajello}}, \citenamefont {{Gasparrini}}, \citenamefont
  {{S{\'a}nchez-Conde}}, \citenamefont {{Zaharijas}}, \citenamefont
  {{Gustafsson}}, \citenamefont {{Cohen-Tanugi}}, \citenamefont {{Dermer}},
  \citenamefont {{Inoue}}, \citenamefont {{Hartmann}}, \citenamefont
  {{Ackermann}}, \citenamefont {{Bechtol}}, \citenamefont {{Franckowiak}},
  \citenamefont {{Reimer}}, \citenamefont {{Romani}},\ and\ \citenamefont
  {{Strong}}}]{2015ApJ...800L..27A}%
  \BibitemOpen
  \bibfield  {author} {\bibinfo {author} {\bibfnamefont {M.}~\bibnamefont
  {{Ajello}}}, \bibinfo {author} {\bibfnamefont {D.}~\bibnamefont
  {{Gasparrini}}}, \bibinfo {author} {\bibfnamefont {M.}~\bibnamefont
  {{S{\'a}nchez-Conde}}}, \bibinfo {author} {\bibfnamefont {G.}~\bibnamefont
  {{Zaharijas}}}, \bibinfo {author} {\bibfnamefont {M.}~\bibnamefont
  {{Gustafsson}}}, \bibinfo {author} {\bibfnamefont {J.}~\bibnamefont
  {{Cohen-Tanugi}}}, \bibinfo {author} {\bibfnamefont {C.~D.}\ \bibnamefont
  {{Dermer}}}, \bibinfo {author} {\bibfnamefont {Y.}~\bibnamefont {{Inoue}}},
  \bibinfo {author} {\bibfnamefont {D.}~\bibnamefont {{Hartmann}}}, \bibinfo
  {author} {\bibfnamefont {M.}~\bibnamefont {{Ackermann}}}, \bibinfo {author}
  {\bibfnamefont {K.}~\bibnamefont {{Bechtol}}}, \bibinfo {author}
  {\bibfnamefont {A.}~\bibnamefont {{Franckowiak}}}, \bibinfo {author}
  {\bibfnamefont {A.}~\bibnamefont {{Reimer}}}, \bibinfo {author}
  {\bibfnamefont {R.~W.}\ \bibnamefont {{Romani}}}, \ and\ \bibinfo {author}
  {\bibfnamefont {A.~W.}\ \bibnamefont {{Strong}}},\ }\href {\doibase
  10.1088/2041-8205/800/2/L27} {\bibfield  {journal} {\bibinfo  {journal}
  {Astrophys.J.Lett.}\ }\textbf {\bibinfo {volume} {800}},\ \bibinfo {eid}
  {L27} (\bibinfo {year} {2015})},\ \Eprint {http://arxiv.org/abs/1501.05301}
  {arXiv:1501.05301 [astro-ph.HE]} \BibitemShut {NoStop}%
\bibitem [{\citenamefont {Fornasa}\ and\ \citenamefont
  {S{\'a}nchez-Conde}(2015)}]{Fornasa:2015qua}%
  \BibitemOpen
  \bibfield  {author} {\bibinfo {author} {\bibfnamefont {M.}~\bibnamefont
  {Fornasa}}\ and\ \bibinfo {author} {\bibfnamefont {M.~A.}\ \bibnamefont
  {S{\'a}nchez-Conde}},\ }\href {\doibase 10.1016/j.physrep.2015.09.002}
  {\bibfield  {journal} {\bibinfo  {journal} {Phys. Rept.}\ }\textbf {\bibinfo
  {volume} {598}},\ \bibinfo {pages} {1} (\bibinfo {year} {2015})},\ \Eprint
  {http://arxiv.org/abs/1502.02866} {arXiv:1502.02866 [astro-ph.CO]}
  \BibitemShut {NoStop}%
\bibitem [{\citenamefont {{Xia}}\ \emph {et~al.}(2015)\citenamefont {{Xia}},
  \citenamefont {{Cuoco}}, \citenamefont {{Branchini}},\ and\ \citenamefont
  {{Viel}}}]{2015ApJS..217...15X}%
  \BibitemOpen
  \bibfield  {author} {\bibinfo {author} {\bibfnamefont {J.-Q.}\ \bibnamefont
  {{Xia}}}, \bibinfo {author} {\bibfnamefont {A.}~\bibnamefont {{Cuoco}}},
  \bibinfo {author} {\bibfnamefont {E.}~\bibnamefont {{Branchini}}}, \ and\
  \bibinfo {author} {\bibfnamefont {M.}~\bibnamefont {{Viel}}},\ }\href
  {\doibase 10.1088/0067-0049/217/1/15} {\bibfield  {journal} {\bibinfo
  {journal} {Astrophys.J.Suppl.}\ }\textbf {\bibinfo {volume} {217}},\ \bibinfo
  {eid} {15} (\bibinfo {year} {2015})},\ \Eprint
  {http://arxiv.org/abs/1503.05918} {arXiv:1503.05918} \BibitemShut {NoStop}%
\bibitem [{\citenamefont {{Fornengo}}\ \emph {et~al.}(2015)\citenamefont
  {{Fornengo}}, \citenamefont {{Perotto}}, \citenamefont {{Regis}},\ and\
  \citenamefont {{Camera}}}]{2015ApJ...802L...1F}%
  \BibitemOpen
  \bibfield  {author} {\bibinfo {author} {\bibfnamefont {N.}~\bibnamefont
  {{Fornengo}}}, \bibinfo {author} {\bibfnamefont {L.}~\bibnamefont
  {{Perotto}}}, \bibinfo {author} {\bibfnamefont {M.}~\bibnamefont {{Regis}}},
  \ and\ \bibinfo {author} {\bibfnamefont {S.}~\bibnamefont {{Camera}}},\
  }\href {\doibase 10.1088/2041-8205/802/1/L1} {\bibfield  {journal} {\bibinfo
  {journal} {Astrophys.J.Lett.}\ }\textbf {\bibinfo {volume} {802}},\ \bibinfo
  {eid} {L1} (\bibinfo {year} {2015})},\ \Eprint
  {http://arxiv.org/abs/1410.4997} {arXiv:1410.4997} \BibitemShut {NoStop}%
\bibitem [{\citenamefont {Cuoco}\ \emph {et~al.}(2015)\citenamefont {Cuoco},
  \citenamefont {Xia}, \citenamefont {Regis}, \citenamefont {Branchini},
  \citenamefont {Fornengo},\ and\ \citenamefont {Viel}}]{Cuoco:2015rfa}%
  \BibitemOpen
  \bibfield  {author} {\bibinfo {author} {\bibfnamefont {A.}~\bibnamefont
  {Cuoco}}, \bibinfo {author} {\bibfnamefont {J.-Q.}\ \bibnamefont {Xia}},
  \bibinfo {author} {\bibfnamefont {M.}~\bibnamefont {Regis}}, \bibinfo
  {author} {\bibfnamefont {E.}~\bibnamefont {Branchini}}, \bibinfo {author}
  {\bibfnamefont {N.}~\bibnamefont {Fornengo}}, \ and\ \bibinfo {author}
  {\bibfnamefont {M.}~\bibnamefont {Viel}},\ }\href@noop {} {\  (\bibinfo
  {year} {2015})},\ \Eprint {http://arxiv.org/abs/1506.01030} {arXiv:1506.01030
  [astro-ph.HE]} \BibitemShut {NoStop}%
\bibitem [{\citenamefont {Gilmore}\ \emph {et~al.}(2012)\citenamefont
  {Gilmore}, \citenamefont {Somerville}, \citenamefont {Primack},\ and\
  \citenamefont {Dominguez}}]{Gilmore:2011ks}%
  \BibitemOpen
  \bibfield  {author} {\bibinfo {author} {\bibfnamefont {R.}~\bibnamefont
  {Gilmore}}, \bibinfo {author} {\bibfnamefont {R.}~\bibnamefont {Somerville}},
  \bibinfo {author} {\bibfnamefont {J.}~\bibnamefont {Primack}}, \ and\
  \bibinfo {author} {\bibfnamefont {A.}~\bibnamefont {Dominguez}},\ }\href@noop
  {} {\bibfield  {journal} {\bibinfo  {journal} {Mon.Not.Roy.Astron.Soc.}\
  }\textbf {\bibinfo {volume} {422}},\ \bibinfo {pages} {3189} (\bibinfo {year}
  {2012})},\ \Eprint {http://arxiv.org/abs/1104.0671} {arXiv:1104.0671
  [astro-ph.CO]} \BibitemShut {NoStop}%
\bibitem [{\citenamefont {Cirelli}\ \emph {et~al.}(2011)\citenamefont
  {Cirelli}, \citenamefont {Corcella}, \citenamefont {Hektor}, \citenamefont
  {Hutsi}, \citenamefont {Kadastik} \emph {et~al.}}]{Cirelli:2010xx}%
  \BibitemOpen
  \bibfield  {author} {\bibinfo {author} {\bibfnamefont {M.}~\bibnamefont
  {Cirelli}}, \bibinfo {author} {\bibfnamefont {G.}~\bibnamefont {Corcella}},
  \bibinfo {author} {\bibfnamefont {A.}~\bibnamefont {Hektor}}, \bibinfo
  {author} {\bibfnamefont {G.}~\bibnamefont {Hutsi}}, \bibinfo {author}
  {\bibfnamefont {M.}~\bibnamefont {Kadastik}},  \emph {et~al.},\ }\href
  {\doibase 10.1088/1475-7516/2012/10/E01, 10.1088/1475-7516/2011/03/051}
  {\bibfield  {journal} {\bibinfo  {journal} {JCAP}\ }\textbf {\bibinfo
  {volume} {1103}},\ \bibinfo {pages} {051} (\bibinfo {year} {2011})},\ \Eprint
  {http://arxiv.org/abs/1012.4515} {arXiv:1012.4515 [hep-ph]} \BibitemShut
  {NoStop}%
\bibitem [{\citenamefont {Tinker}\ \emph {et~al.}(2008)\citenamefont {Tinker},
  \citenamefont {Kravtsov}, \citenamefont {Klypin}, \citenamefont {Abazajian},
  \citenamefont {Warren} \emph {et~al.}}]{Tinker:2008ff}%
  \BibitemOpen
  \bibfield  {author} {\bibinfo {author} {\bibfnamefont {J.~L.}\ \bibnamefont
  {Tinker}}, \bibinfo {author} {\bibfnamefont {A.~V.}\ \bibnamefont
  {Kravtsov}}, \bibinfo {author} {\bibfnamefont {A.}~\bibnamefont {Klypin}},
  \bibinfo {author} {\bibfnamefont {K.}~\bibnamefont {Abazajian}}, \bibinfo
  {author} {\bibfnamefont {M.~S.}\ \bibnamefont {Warren}},  \emph {et~al.},\
  }\href {\doibase 10.1086/591439} {\bibfield  {journal} {\bibinfo  {journal}
  {Astrophys.J.}\ }\textbf {\bibinfo {volume} {688}},\ \bibinfo {pages} {709}
  (\bibinfo {year} {2008})},\ \Eprint {http://arxiv.org/abs/0803.2706}
  {arXiv:0803.2706 [astro-ph]} \BibitemShut {NoStop}%
\bibitem [{\citenamefont {Tinker}\ \emph {et~al.}(2010)\citenamefont {Tinker},
  \citenamefont {Robertson}, \citenamefont {Kravtsov}, \citenamefont {Klypin},
  \citenamefont {Warren} \emph {et~al.}}]{Tinker:2010my}%
  \BibitemOpen
  \bibfield  {author} {\bibinfo {author} {\bibfnamefont {J.~L.}\ \bibnamefont
  {Tinker}}, \bibinfo {author} {\bibfnamefont {B.~E.}\ \bibnamefont
  {Robertson}}, \bibinfo {author} {\bibfnamefont {A.~V.}\ \bibnamefont
  {Kravtsov}}, \bibinfo {author} {\bibfnamefont {A.}~\bibnamefont {Klypin}},
  \bibinfo {author} {\bibfnamefont {M.~S.}\ \bibnamefont {Warren}},  \emph
  {et~al.},\ }\href {\doibase 10.1088/0004-637X/724/2/878} {\bibfield
  {journal} {\bibinfo  {journal} {Astrophys.J.}\ }\textbf {\bibinfo {volume}
  {724}},\ \bibinfo {pages} {878} (\bibinfo {year} {2010})},\ \Eprint
  {http://arxiv.org/abs/1001.3162} {arXiv:1001.3162 [astro-ph.CO]} \BibitemShut
  {NoStop}%
\bibitem [{\citenamefont {Ando}\ and\ \citenamefont
  {Komatsu}(2013)}]{Ando:2013ff}%
  \BibitemOpen
  \bibfield  {author} {\bibinfo {author} {\bibfnamefont {S.}~\bibnamefont
  {Ando}}\ and\ \bibinfo {author} {\bibfnamefont {E.}~\bibnamefont {Komatsu}},\
  }\href {\doibase 10.1103/PhysRevD.87.123539} {\bibfield  {journal} {\bibinfo
  {journal} {Phys.Rev.}\ }\textbf {\bibinfo {volume} {D87}},\ \bibinfo {pages}
  {123539} (\bibinfo {year} {2013})},\ \Eprint {http://arxiv.org/abs/1301.5901}
  {arXiv:1301.5901 [astro-ph.CO]} \BibitemShut {NoStop}%
\bibitem [{\citenamefont {Navarro}\ \emph {et~al.}(1997)\citenamefont
  {Navarro}, \citenamefont {Frenk},\ and\ \citenamefont
  {White}}]{Navarro:1996gj}%
  \BibitemOpen
  \bibfield  {author} {\bibinfo {author} {\bibfnamefont {J.~F.}\ \bibnamefont
  {Navarro}}, \bibinfo {author} {\bibfnamefont {C.~S.}\ \bibnamefont {Frenk}},
  \ and\ \bibinfo {author} {\bibfnamefont {S.~D.}\ \bibnamefont {White}},\
  }\href {\doibase 10.1086/304888} {\bibfield  {journal} {\bibinfo  {journal}
  {Astrophys.J.}\ }\textbf {\bibinfo {volume} {490}},\ \bibinfo {pages} {493}
  (\bibinfo {year} {1997})},\ \Eprint {http://arxiv.org/abs/astro-ph/9611107}
  {arXiv:astro-ph/9611107 [astro-ph]} \BibitemShut {NoStop}%
\bibitem [{\citenamefont {Prada}\ \emph {et~al.}(2012)\citenamefont {Prada},
  \citenamefont {Klypin}, \citenamefont {Cuesta}, \citenamefont
  {Betancort-Rijo},\ and\ \citenamefont {Primack}}]{Prada:2011jf}%
  \BibitemOpen
  \bibfield  {author} {\bibinfo {author} {\bibfnamefont {F.}~\bibnamefont
  {Prada}}, \bibinfo {author} {\bibfnamefont {A.~A.}\ \bibnamefont {Klypin}},
  \bibinfo {author} {\bibfnamefont {A.~J.}\ \bibnamefont {Cuesta}}, \bibinfo
  {author} {\bibfnamefont {J.~E.}\ \bibnamefont {Betancort-Rijo}}, \ and\
  \bibinfo {author} {\bibfnamefont {J.}~\bibnamefont {Primack}},\ }\href
  {\doibase 10.1111/j.1365-2966.2012.21007.x} {\bibfield  {journal} {\bibinfo
  {journal} {Mon. Not. Roy. Astron. Soc.}\ }\textbf {\bibinfo {volume} {428}},\
  \bibinfo {pages} {3018} (\bibinfo {year} {2012})},\ \Eprint
  {http://arxiv.org/abs/1104.5130} {arXiv:1104.5130 [astro-ph.CO]} \BibitemShut
  {NoStop}%
\bibitem [{\citenamefont {Gao}\ \emph {et~al.}(2012)\citenamefont {Gao},
  \citenamefont {Frenk}, \citenamefont {Jenkins}, \citenamefont {Springel},\
  and\ \citenamefont {White}}]{Gao:2011rf}%
  \BibitemOpen
  \bibfield  {author} {\bibinfo {author} {\bibfnamefont {L.}~\bibnamefont
  {Gao}}, \bibinfo {author} {\bibfnamefont {C.}~\bibnamefont {Frenk}}, \bibinfo
  {author} {\bibfnamefont {A.}~\bibnamefont {Jenkins}}, \bibinfo {author}
  {\bibfnamefont {V.}~\bibnamefont {Springel}}, \ and\ \bibinfo {author}
  {\bibfnamefont {S.}~\bibnamefont {White}},\ }\href {\doibase
  10.1111/j.1365-2966.2011.19836.x} {\bibfield  {journal} {\bibinfo  {journal}
  {Mon.Not.Roy.Astron.Soc.}\ }\textbf {\bibinfo {volume} {419}},\ \bibinfo
  {pages} {1721} (\bibinfo {year} {2012})},\ \Eprint
  {http://arxiv.org/abs/1107.1916} {arXiv:1107.1916 [astro-ph.CO]} \BibitemShut
  {NoStop}%
\bibitem [{\citenamefont {S{\'a}nchez-Conde}\ and\ \citenamefont
  {Prada}(2014)}]{Sanchez-Conde:2013yxa}%
  \BibitemOpen
  \bibfield  {author} {\bibinfo {author} {\bibfnamefont {M.~A.}\ \bibnamefont
  {S{\'a}nchez-Conde}}\ and\ \bibinfo {author} {\bibfnamefont {F.}~\bibnamefont
  {Prada}},\ }\href {\doibase 10.1093/mnras/stu1014} {\bibfield  {journal}
  {\bibinfo  {journal} {Mon. Not. Roy. Astron. Soc.}\ }\textbf {\bibinfo
  {volume} {442}},\ \bibinfo {pages} {2271} (\bibinfo {year} {2014})},\ \Eprint
  {http://arxiv.org/abs/1312.1729} {arXiv:1312.1729 [astro-ph.CO]} \BibitemShut
  {NoStop}%
\bibitem [{\citenamefont {Bartels}\ and\ \citenamefont
  {Ando}(2015)}]{Bartels:2015uba}%
  \BibitemOpen
  \bibfield  {author} {\bibinfo {author} {\bibfnamefont {R.}~\bibnamefont
  {Bartels}}\ and\ \bibinfo {author} {\bibfnamefont {S.}~\bibnamefont {Ando}},\
  }\href@noop {} {\  (\bibinfo {year} {2015})},\ \Eprint
  {http://arxiv.org/abs/1507.08656} {arXiv:1507.08656 [astro-ph.CO]}
  \BibitemShut {NoStop}%
\bibitem [{\citenamefont {Reid}\ and\ \citenamefont
  {Spergel}(2009)}]{Reid:2008sy}%
  \BibitemOpen
  \bibfield  {author} {\bibinfo {author} {\bibfnamefont {B.~A.}\ \bibnamefont
  {Reid}}\ and\ \bibinfo {author} {\bibfnamefont {D.~N.}\ \bibnamefont
  {Spergel}},\ }\href {\doibase 10.1088/0004-637X/698/1/143} {\bibfield
  {journal} {\bibinfo  {journal} {Astrophys.J.}\ }\textbf {\bibinfo {volume}
  {698}},\ \bibinfo {pages} {143} (\bibinfo {year} {2009})},\ \Eprint
  {http://arxiv.org/abs/0809.4505} {arXiv:0809.4505 [astro-ph]} \BibitemShut
  {NoStop}%
\bibitem [{\citenamefont {Wake}\ \emph {et~al.}(2008)\citenamefont {Wake},
  \citenamefont {Sheth}, \citenamefont {Nichol}, \citenamefont {Baugh},
  \citenamefont {Bland-Hawthorn} \emph {et~al.}}]{Wake:2008mf}%
  \BibitemOpen
  \bibfield  {author} {\bibinfo {author} {\bibfnamefont {D.~A.}\ \bibnamefont
  {Wake}}, \bibinfo {author} {\bibfnamefont {R.~K.}\ \bibnamefont {Sheth}},
  \bibinfo {author} {\bibfnamefont {R.~C.}\ \bibnamefont {Nichol}}, \bibinfo
  {author} {\bibfnamefont {C.~M.}\ \bibnamefont {Baugh}}, \bibinfo {author}
  {\bibfnamefont {J.}~\bibnamefont {Bland-Hawthorn}},  \emph {et~al.},\ }\href
  {\doibase 10.1111/j.1365-2966.2008.13333.x} {\bibfield  {journal} {\bibinfo
  {journal} {Mon.Not.Roy.Astron.Soc.}\ }\textbf {\bibinfo {volume} {387}},\
  \bibinfo {pages} {1045} (\bibinfo {year} {2008})},\ \Eprint
  {http://arxiv.org/abs/0802.4288} {arXiv:0802.4288 [astro-ph]} \BibitemShut
  {NoStop}%
\bibitem [{\citenamefont {Hikage}\ \emph {et~al.}(2013)\citenamefont {Hikage},
  \citenamefont {Mandelbaum}, \citenamefont {Takada},\ and\ \citenamefont
  {Spergel}}]{Hikage:2012zk}%
  \BibitemOpen
  \bibfield  {author} {\bibinfo {author} {\bibfnamefont {C.}~\bibnamefont
  {Hikage}}, \bibinfo {author} {\bibfnamefont {R.}~\bibnamefont {Mandelbaum}},
  \bibinfo {author} {\bibfnamefont {M.}~\bibnamefont {Takada}}, \ and\ \bibinfo
  {author} {\bibfnamefont {D.~N.}\ \bibnamefont {Spergel}},\ }\href {\doibase
  10.1093/mnras/stt1446} {\bibfield  {journal} {\bibinfo  {journal}
  {Mon.Not.Roy.Astron.Soc.}\ }\textbf {\bibinfo {volume} {435}},\ \bibinfo
  {pages} {2345} (\bibinfo {year} {2013})},\ \Eprint
  {http://arxiv.org/abs/1211.1009} {arXiv:1211.1009 [astro-ph.CO]} \BibitemShut
  {NoStop}%
\bibitem [{\citenamefont {Limber}(1954)}]{Limber:1954zz}%
  \BibitemOpen
  \bibfield  {author} {\bibinfo {author} {\bibfnamefont {D.~N.}\ \bibnamefont
  {Limber}},\ }\href@noop {} {\bibfield  {journal} {\bibinfo  {journal}
  {Astrophys.J.}\ }\textbf {\bibinfo {volume} {119}},\ \bibinfo {pages} {655}
  (\bibinfo {year} {1954})}\BibitemShut {NoStop}%
\bibitem [{\citenamefont {Fukugita}\ \emph {et~al.}(1996)\citenamefont
  {Fukugita}, \citenamefont {Ichikawa}, \citenamefont {Gunn}, \citenamefont
  {Doi}, \citenamefont {Shimasaku} \emph {et~al.}}]{Fukugita:1996qt}%
  \BibitemOpen
  \bibfield  {author} {\bibinfo {author} {\bibfnamefont {M.}~\bibnamefont
  {Fukugita}}, \bibinfo {author} {\bibfnamefont {T.}~\bibnamefont {Ichikawa}},
  \bibinfo {author} {\bibfnamefont {J.}~\bibnamefont {Gunn}}, \bibinfo {author}
  {\bibfnamefont {M.}~\bibnamefont {Doi}}, \bibinfo {author} {\bibfnamefont
  {K.}~\bibnamefont {Shimasaku}},  \emph {et~al.},\ }\href {\doibase
  10.1086/117915} {\bibfield  {journal} {\bibinfo  {journal} {Astron.J.}\
  }\textbf {\bibinfo {volume} {111}},\ \bibinfo {pages} {1748} (\bibinfo {year}
  {1996})}\BibitemShut {NoStop}%
\bibitem [{\citenamefont {Gunn}\ \emph {et~al.}(1998)\citenamefont {Gunn} \emph
  {et~al.}}]{Gunn:1998vh}%
  \BibitemOpen
  \bibfield  {author} {\bibinfo {author} {\bibfnamefont {J.}~\bibnamefont
  {Gunn}} \emph {et~al.} (\bibinfo {collaboration} {SDSS}),\ }\href {\doibase
  10.1086/300645} {\bibfield  {journal} {\bibinfo  {journal} {Astron.J.}\
  }\textbf {\bibinfo {volume} {116}},\ \bibinfo {pages} {3040} (\bibinfo {year}
  {1998})},\ \Eprint {http://arxiv.org/abs/astro-ph/9809085}
  {arXiv:astro-ph/9809085 [astro-ph]} \BibitemShut {NoStop}%
\bibitem [{\citenamefont {Gunn}\ \emph {et~al.}(2006)\citenamefont {Gunn} \emph
  {et~al.}}]{Gunn:2006tw}%
  \BibitemOpen
  \bibfield  {author} {\bibinfo {author} {\bibfnamefont {J.~E.}\ \bibnamefont
  {Gunn}} \emph {et~al.} (\bibinfo {collaboration} {SDSS}),\ }\href {\doibase
  10.1086/500975} {\bibfield  {journal} {\bibinfo  {journal} {Astron.J.}\
  }\textbf {\bibinfo {volume} {131}},\ \bibinfo {pages} {2332} (\bibinfo {year}
  {2006})},\ \Eprint {http://arxiv.org/abs/astro-ph/0602326}
  {arXiv:astro-ph/0602326 [astro-ph]} \BibitemShut {NoStop}%
\bibitem [{\citenamefont {Lupton}\ \emph {et~al.}(2001)\citenamefont {Lupton}
  \emph {et~al.}}]{Lupton:2001zb}%
  \BibitemOpen
  \bibfield  {author} {\bibinfo {author} {\bibfnamefont {R.}~\bibnamefont
  {Lupton}} \emph {et~al.} (\bibinfo {collaboration} {SDSS}),\ }\href@noop {}
  {\bibfield  {journal} {\bibinfo  {journal} {ASP Conf.Ser.}\ }\textbf
  {\bibinfo {volume} {238}},\ \bibinfo {pages} {269} (\bibinfo {year}
  {2001})},\ \Eprint {http://arxiv.org/abs/astro-ph/0101420}
  {arXiv:astro-ph/0101420 [astro-ph]} \BibitemShut {NoStop}%
\bibitem [{\citenamefont {Stoughton}\ \emph {et~al.}(2002)\citenamefont
  {Stoughton} \emph {et~al.}}]{Stoughton:2002ae}%
  \BibitemOpen
  \bibfield  {author} {\bibinfo {author} {\bibfnamefont {C.}~\bibnamefont
  {Stoughton}} \emph {et~al.} (\bibinfo {collaboration} {SDSS}),\ }\href
  {\doibase 10.1086/324741} {\bibfield  {journal} {\bibinfo  {journal}
  {Astron.J.}\ }\textbf {\bibinfo {volume} {123}},\ \bibinfo {pages} {485}
  (\bibinfo {year} {2002})}\BibitemShut {NoStop}%
\bibitem [{\citenamefont {Pier}\ \emph {et~al.}(2003)\citenamefont {Pier},
  \citenamefont {Munn}, \citenamefont {Hindsley}, \citenamefont {Hennessy},
  \citenamefont {Kent} \emph {et~al.}}]{Pier:2002iq}%
  \BibitemOpen
  \bibfield  {author} {\bibinfo {author} {\bibfnamefont {J.~R.}\ \bibnamefont
  {Pier}}, \bibinfo {author} {\bibfnamefont {J.~A.}\ \bibnamefont {Munn}},
  \bibinfo {author} {\bibfnamefont {R.~B.}\ \bibnamefont {Hindsley}}, \bibinfo
  {author} {\bibfnamefont {G.}~\bibnamefont {Hennessy}}, \bibinfo {author}
  {\bibfnamefont {S.~M.}\ \bibnamefont {Kent}},  \emph {et~al.},\ }\href
  {\doibase 10.1086/346138} {\bibfield  {journal} {\bibinfo  {journal}
  {Astron.J.}\ }\textbf {\bibinfo {volume} {125}},\ \bibinfo {pages} {1559}
  (\bibinfo {year} {2003})},\ \Eprint {http://arxiv.org/abs/astro-ph/0211375}
  {arXiv:astro-ph/0211375 [astro-ph]} \BibitemShut {NoStop}%
\bibitem [{\citenamefont {Ivezic}\ \emph {et~al.}(2004)\citenamefont {Ivezic}
  \emph {et~al.}}]{Ivezic:2004bf}%
  \BibitemOpen
  \bibfield  {author} {\bibinfo {author} {\bibfnamefont {Z.}~\bibnamefont
  {Ivezic}} \emph {et~al.} (\bibinfo {collaboration} {SDSS}),\ }\href {\doibase
  10.1002/asna.200410285} {\bibfield  {journal} {\bibinfo  {journal}
  {Astron.Nachr.}\ }\textbf {\bibinfo {volume} {325}},\ \bibinfo {pages} {583}
  (\bibinfo {year} {2004})},\ \Eprint {http://arxiv.org/abs/astro-ph/0410195}
  {arXiv:astro-ph/0410195 [astro-ph]} \BibitemShut {NoStop}%
\bibitem [{\citenamefont {Hogg}\ \emph {et~al.}(2001)\citenamefont {Hogg},
  \citenamefont {Finkbeiner}, \citenamefont {Schlegel},\ and\ \citenamefont
  {Gunn}}]{Hogg:2001gc}%
  \BibitemOpen
  \bibfield  {author} {\bibinfo {author} {\bibfnamefont {D.~W.}\ \bibnamefont
  {Hogg}}, \bibinfo {author} {\bibfnamefont {D.~P.}\ \bibnamefont
  {Finkbeiner}}, \bibinfo {author} {\bibfnamefont {D.~J.}\ \bibnamefont
  {Schlegel}}, \ and\ \bibinfo {author} {\bibfnamefont {J.~E.}\ \bibnamefont
  {Gunn}},\ }\href {\doibase 10.1086/323103} {\bibfield  {journal} {\bibinfo
  {journal} {Astron.J.}\ }\textbf {\bibinfo {volume} {122}},\ \bibinfo {pages}
  {2129} (\bibinfo {year} {2001})},\ \Eprint
  {http://arxiv.org/abs/astro-ph/0106511} {arXiv:astro-ph/0106511 [astro-ph]}
  \BibitemShut {NoStop}%
\bibitem [{\citenamefont {Smith}\ \emph {et~al.}(2002)\citenamefont {Smith}
  \emph {et~al.}}]{Smith:2002pca}%
  \BibitemOpen
  \bibfield  {author} {\bibinfo {author} {\bibfnamefont {J.~A.}\ \bibnamefont
  {Smith}} \emph {et~al.} (\bibinfo {collaboration} {SDSS}),\ }\href {\doibase
  10.1086/339311} {\bibfield  {journal} {\bibinfo  {journal} {Astron.J.}\
  }\textbf {\bibinfo {volume} {123}},\ \bibinfo {pages} {2121} (\bibinfo {year}
  {2002})},\ \Eprint {http://arxiv.org/abs/astro-ph/0201143}
  {arXiv:astro-ph/0201143 [astro-ph]} \BibitemShut {NoStop}%
\bibitem [{\citenamefont {Tucker}\ \emph {et~al.}(2006)\citenamefont {Tucker}
  \emph {et~al.}}]{Tucker:2006dv}%
  \BibitemOpen
  \bibfield  {author} {\bibinfo {author} {\bibfnamefont {D.}~\bibnamefont
  {Tucker}} \emph {et~al.} (\bibinfo {collaboration} {SDSS}),\ }\href {\doibase
  10.1002/asna.200610655} {\bibfield  {journal} {\bibinfo  {journal}
  {Astron.Nachr.}\ }\textbf {\bibinfo {volume} {327}},\ \bibinfo {pages} {821}
  (\bibinfo {year} {2006})},\ \Eprint {http://arxiv.org/abs/astro-ph/0608575}
  {arXiv:astro-ph/0608575 [astro-ph]} \BibitemShut {NoStop}%
\bibitem [{\citenamefont {Eisenstein}\ \emph {et~al.}(2001)\citenamefont
  {Eisenstein} \emph {et~al.}}]{Eisenstein:2001cq}%
  \BibitemOpen
  \bibfield  {author} {\bibinfo {author} {\bibfnamefont {D.~J.}\ \bibnamefont
  {Eisenstein}} \emph {et~al.} (\bibinfo {collaboration} {SDSS}),\ }\href
  {\doibase 10.1086/323717} {\bibfield  {journal} {\bibinfo  {journal}
  {Astron.J.}\ }\textbf {\bibinfo {volume} {122}},\ \bibinfo {pages} {2267}
  (\bibinfo {year} {2001})},\ \Eprint {http://arxiv.org/abs/astro-ph/0108153}
  {arXiv:astro-ph/0108153 [astro-ph]} \BibitemShut {NoStop}%
\bibitem [{\citenamefont {Strauss}\ \emph {et~al.}(2002)\citenamefont {Strauss}
  \emph {et~al.}}]{Strauss:2002dj}%
  \BibitemOpen
  \bibfield  {author} {\bibinfo {author} {\bibfnamefont {M.~A.}\ \bibnamefont
  {Strauss}} \emph {et~al.} (\bibinfo {collaboration} {SDSS}),\ }\href
  {\doibase 10.1086/342343} {\bibfield  {journal} {\bibinfo  {journal}
  {Astron.J.}\ }\textbf {\bibinfo {volume} {124}},\ \bibinfo {pages} {1810}
  (\bibinfo {year} {2002})},\ \Eprint {http://arxiv.org/abs/astro-ph/0206225}
  {arXiv:astro-ph/0206225 [astro-ph]} \BibitemShut {NoStop}%
\bibitem [{\citenamefont {Richards}\ \emph {et~al.}(2002)\citenamefont
  {Richards} \emph {et~al.}}]{Richards:2002bb}%
  \BibitemOpen
  \bibfield  {author} {\bibinfo {author} {\bibfnamefont {G.~T.}\ \bibnamefont
  {Richards}} \emph {et~al.} (\bibinfo {collaboration} {SDSS}),\ }\href
  {\doibase 10.1086/340187} {\bibfield  {journal} {\bibinfo  {journal}
  {Astron.J.}\ }\textbf {\bibinfo {volume} {123}},\ \bibinfo {pages} {2945}
  (\bibinfo {year} {2002})},\ \Eprint {http://arxiv.org/abs/astro-ph/0202251}
  {arXiv:astro-ph/0202251 [astro-ph]} \BibitemShut {NoStop}%
\bibitem [{\citenamefont {Blanton}\ \emph {et~al.}(2003)\citenamefont
  {Blanton}, \citenamefont {Lupton}, \citenamefont {Miller~Malley},
  \citenamefont {Young}, \citenamefont {Zehavi} \emph
  {et~al.}}]{Blanton:2001yk}%
  \BibitemOpen
  \bibfield  {author} {\bibinfo {author} {\bibfnamefont {M.~R.}\ \bibnamefont
  {Blanton}}, \bibinfo {author} {\bibfnamefont {R.~H.}\ \bibnamefont {Lupton}},
  \bibinfo {author} {\bibfnamefont {F.}~\bibnamefont {Miller~Malley}}, \bibinfo
  {author} {\bibfnamefont {N.}~\bibnamefont {Young}}, \bibinfo {author}
  {\bibfnamefont {I.}~\bibnamefont {Zehavi}},  \emph {et~al.},\ }\href
  {\doibase 10.1086/344761} {\bibfield  {journal} {\bibinfo  {journal}
  {Astron.J.}\ }\textbf {\bibinfo {volume} {125}},\ \bibinfo {pages} {2276}
  (\bibinfo {year} {2003})},\ \Eprint {http://arxiv.org/abs/astro-ph/0105535}
  {arXiv:astro-ph/0105535 [astro-ph]} \BibitemShut {NoStop}%
\bibitem [{\citenamefont {Abazajian}\ \emph {et~al.}(2009)\citenamefont
  {Abazajian} \emph {et~al.}}]{Abazajian:2008wr}%
  \BibitemOpen
  \bibfield  {author} {\bibinfo {author} {\bibfnamefont {K.~N.}\ \bibnamefont
  {Abazajian}} \emph {et~al.} (\bibinfo {collaboration} {SDSS}),\ }\href
  {\doibase 10.1088/0067-0049/182/2/543} {\bibfield  {journal} {\bibinfo
  {journal} {Astrophys.J.Suppl.}\ }\textbf {\bibinfo {volume} {182}},\ \bibinfo
  {pages} {543} (\bibinfo {year} {2009})},\ \Eprint
  {http://arxiv.org/abs/0812.0649} {arXiv:0812.0649 [astro-ph]} \BibitemShut
  {NoStop}%
\bibitem [{\citenamefont {Kazin}\ \emph {et~al.}(2010)\citenamefont {Kazin}
  \emph {et~al.}}]{Kazin:2009cj}%
  \BibitemOpen
  \bibfield  {author} {\bibinfo {author} {\bibfnamefont {E.~A.}\ \bibnamefont
  {Kazin}} \emph {et~al.} (\bibinfo {collaboration} {SDSS}),\ }\href {\doibase
  10.1088/0004-637X/710/2/1444} {\bibfield  {journal} {\bibinfo  {journal}
  {Astrophys.J.}\ }\textbf {\bibinfo {volume} {710}},\ \bibinfo {pages} {1444}
  (\bibinfo {year} {2010})},\ \Eprint {http://arxiv.org/abs/0908.2598}
  {arXiv:0908.2598 [astro-ph.CO]} \BibitemShut {NoStop}%
\bibitem [{\citenamefont {Xia}\ \emph {et~al.}(2011)\citenamefont {Xia},
  \citenamefont {Cuoco}, \citenamefont {Branchini}, \citenamefont {Fornasa},\
  and\ \citenamefont {Viel}}]{Xia:2011ax}%
  \BibitemOpen
  \bibfield  {author} {\bibinfo {author} {\bibfnamefont {J.-Q.}\ \bibnamefont
  {Xia}}, \bibinfo {author} {\bibfnamefont {A.}~\bibnamefont {Cuoco}}, \bibinfo
  {author} {\bibfnamefont {E.}~\bibnamefont {Branchini}}, \bibinfo {author}
  {\bibfnamefont {M.}~\bibnamefont {Fornasa}}, \ and\ \bibinfo {author}
  {\bibfnamefont {M.}~\bibnamefont {Viel}},\ }\href@noop {} {\bibfield
  {journal} {\bibinfo  {journal} {Mon.Not.Roy.Astron.Soc.}\ }\textbf {\bibinfo
  {volume} {416}},\ \bibinfo {pages} {2247} (\bibinfo {year} {2011})},\ \Eprint
  {http://arxiv.org/abs/1103.4861} {arXiv:1103.4861 [astro-ph.CO]} \BibitemShut
  {NoStop}%
\bibitem [{\citenamefont {Su}\ \emph {et~al.}(2010)\citenamefont {Su},
  \citenamefont {Slatyer},\ and\ \citenamefont {Finkbeiner}}]{Su:2010qj}%
  \BibitemOpen
  \bibfield  {author} {\bibinfo {author} {\bibfnamefont {M.}~\bibnamefont
  {Su}}, \bibinfo {author} {\bibfnamefont {T.~R.}\ \bibnamefont {Slatyer}}, \
  and\ \bibinfo {author} {\bibfnamefont {D.~P.}\ \bibnamefont {Finkbeiner}},\
  }\href {\doibase 10.1088/0004-637X/724/2/1044} {\bibfield  {journal}
  {\bibinfo  {journal} {Astrophys.J.}\ }\textbf {\bibinfo {volume} {724}},\
  \bibinfo {pages} {1044} (\bibinfo {year} {2010})},\ \Eprint
  {http://arxiv.org/abs/1005.5480} {arXiv:1005.5480 [astro-ph.HE]} \BibitemShut
  {NoStop}%
\bibitem [{\citenamefont {Shirasaki}\ \emph {et~al.}(2014)\citenamefont
  {Shirasaki}, \citenamefont {Horiuchi},\ and\ \citenamefont
  {Yoshida}}]{Shirasaki:2014noa}%
  \BibitemOpen
  \bibfield  {author} {\bibinfo {author} {\bibfnamefont {M.}~\bibnamefont
  {Shirasaki}}, \bibinfo {author} {\bibfnamefont {S.}~\bibnamefont {Horiuchi}},
  \ and\ \bibinfo {author} {\bibfnamefont {N.}~\bibnamefont {Yoshida}},\ }\href
  {\doibase 10.1103/PhysRevD.90.063502} {\bibfield  {journal} {\bibinfo
  {journal} {Phys.Rev.}\ }\textbf {\bibinfo {volume} {D90}},\ \bibinfo {pages}
  {063502} (\bibinfo {year} {2014})},\ \Eprint {http://arxiv.org/abs/1404.5503}
  {arXiv:1404.5503 [astro-ph.CO]} \BibitemShut {NoStop}%
\bibitem [{\citenamefont {Ackermann}\ \emph
  {et~al.}(2012{\natexlab{a}})\citenamefont {Ackermann} \emph
  {et~al.}}]{Ackermann:2012pya}%
  \BibitemOpen
  \bibfield  {author} {\bibinfo {author} {\bibfnamefont {M.}~\bibnamefont
  {Ackermann}} \emph {et~al.} (\bibinfo {collaboration} {Fermi-LAT}),\ }\href
  {\doibase 10.1088/0004-637X/750/1/3} {\bibfield  {journal} {\bibinfo
  {journal} {Astrophys. J.}\ }\textbf {\bibinfo {volume} {750}},\ \bibinfo
  {pages} {3} (\bibinfo {year} {2012}{\natexlab{a}})},\ \Eprint
  {http://arxiv.org/abs/1202.4039} {arXiv:1202.4039 [astro-ph.HE]} \BibitemShut
  {NoStop}%
\bibitem [{\citenamefont {{Steigman}}\ \emph {et~al.}(2012)\citenamefont
  {{Steigman}}, \citenamefont {{Dasgupta}},\ and\ \citenamefont
  {{Beacom}}}]{2012PhRvD..86b3506S}%
  \BibitemOpen
  \bibfield  {author} {\bibinfo {author} {\bibfnamefont {G.}~\bibnamefont
  {{Steigman}}}, \bibinfo {author} {\bibfnamefont {B.}~\bibnamefont
  {{Dasgupta}}}, \ and\ \bibinfo {author} {\bibfnamefont {J.~F.}\ \bibnamefont
  {{Beacom}}},\ }\href {\doibase 10.1103/PhysRevD.86.023506} {\bibfield
  {journal} {\bibinfo  {journal} {\prd}\ }\textbf {\bibinfo {volume} {86}},\
  \bibinfo {eid} {023506} (\bibinfo {year} {2012})},\ \Eprint
  {http://arxiv.org/abs/1204.3622} {arXiv:1204.3622 [hep-ph]} \BibitemShut
  {NoStop}%
\bibitem [{\citenamefont {Ackermann}\ \emph
  {et~al.}(2012{\natexlab{b}})\citenamefont {Ackermann} \emph
  {et~al.}}]{FermiLAT:2012aa}%
  \BibitemOpen
  \bibfield  {author} {\bibinfo {author} {\bibfnamefont {M.}~\bibnamefont
  {Ackermann}} \emph {et~al.} (\bibinfo {collaboration} {Fermi-LAT}),\ }\href
  {\doibase 10.1088/0004-637X/750/1/3} {\bibfield  {journal} {\bibinfo
  {journal} {Astrophys.J.}\ }\textbf {\bibinfo {volume} {750}},\ \bibinfo
  {pages} {3} (\bibinfo {year} {2012}{\natexlab{b}})},\ \Eprint
  {http://arxiv.org/abs/1202.4039} {arXiv:1202.4039 [astro-ph.HE]} \BibitemShut
  {NoStop}%
\bibitem [{\citenamefont {{Strong}}\ \emph {et~al.}(1999)\citenamefont
  {{Strong}}, \citenamefont {{Moskalenko}},\ and\ \citenamefont
  {{Reimer}}}]{1999ICRC....4...52S}%
  \BibitemOpen
  \bibfield  {author} {\bibinfo {author} {\bibfnamefont {A.}~\bibnamefont
  {{Strong}}}, \bibinfo {author} {\bibfnamefont {I.~V.}\ \bibnamefont
  {{Moskalenko}}}, \ and\ \bibinfo {author} {\bibfnamefont {O.}~\bibnamefont
  {{Reimer}}},\ }\href@noop {} {\bibfield  {journal} {\bibinfo  {journal}
  {International Cosmic Ray Conference}\ }\textbf {\bibinfo {volume} {4}},\
  \bibinfo {pages} {52} (\bibinfo {year} {1999})},\ \Eprint
  {http://arxiv.org/abs/astro-ph/9906229} {astro-ph/9906229} \BibitemShut
  {NoStop}%
\bibitem [{\citenamefont {{Strong}}\ and\ \citenamefont
  {{Moskalenko}}(1999)}]{1999ICRC....4..255S}%
  \BibitemOpen
  \bibfield  {author} {\bibinfo {author} {\bibfnamefont {A.~W.}\ \bibnamefont
  {{Strong}}}\ and\ \bibinfo {author} {\bibfnamefont {I.~V.}\ \bibnamefont
  {{Moskalenko}}},\ }\href@noop {} {\bibfield  {journal} {\bibinfo  {journal}
  {International Cosmic Ray Conference}\ }\textbf {\bibinfo {volume} {4}},\
  \bibinfo {pages} {255} (\bibinfo {year} {1999})},\ \Eprint
  {http://arxiv.org/abs/astro-ph/9906228} {astro-ph/9906228} \BibitemShut
  {NoStop}%
\bibitem [{\citenamefont {{Moskalenko}}\ and\ \citenamefont
  {{Strong}}(2000)}]{2000Ap&SS.272..247M}%
  \BibitemOpen
  \bibfield  {author} {\bibinfo {author} {\bibfnamefont {I.~V.}\ \bibnamefont
  {{Moskalenko}}}\ and\ \bibinfo {author} {\bibfnamefont {A.~W.}\ \bibnamefont
  {{Strong}}},\ }\href {\doibase 10.1023/A:1002604831398} {\bibfield  {journal}
  {\bibinfo  {journal} {Astrophysics and Space Science}\ }\textbf {\bibinfo
  {volume} {272}},\ \bibinfo {pages} {247} (\bibinfo {year} {2000})},\ \Eprint
  {http://arxiv.org/abs/astro-ph/9908032} {astro-ph/9908032} \BibitemShut
  {NoStop}%
\bibitem [{\citenamefont {Case}\ and\ \citenamefont
  {Bhattacharya}(1998)}]{Case:1998qg}%
  \BibitemOpen
  \bibfield  {author} {\bibinfo {author} {\bibfnamefont {G.~L.}\ \bibnamefont
  {Case}}\ and\ \bibinfo {author} {\bibfnamefont {D.}~\bibnamefont
  {Bhattacharya}},\ }\href {\doibase 10.1086/306089} {\bibfield  {journal}
  {\bibinfo  {journal} {Astrophys. J.}\ }\textbf {\bibinfo {volume} {504}},\
  \bibinfo {pages} {761} (\bibinfo {year} {1998})},\ \Eprint
  {http://arxiv.org/abs/astro-ph/9807162} {arXiv:astro-ph/9807162 [astro-ph]}
  \BibitemShut {NoStop}%
\bibitem [{\citenamefont {Bronfman}\ \emph {et~al.}(2000)\citenamefont
  {Bronfman}, \citenamefont {Casassus}, \citenamefont {May},\ and\
  \citenamefont {Nyman}}]{Bronfman:2000tw}%
  \BibitemOpen
  \bibfield  {author} {\bibinfo {author} {\bibfnamefont {L.}~\bibnamefont
  {Bronfman}}, \bibinfo {author} {\bibfnamefont {S.}~\bibnamefont {Casassus}},
  \bibinfo {author} {\bibfnamefont {J.}~\bibnamefont {May}}, \ and\ \bibinfo
  {author} {\bibfnamefont {L.~A.}\ \bibnamefont {Nyman}},\ }\href@noop {}
  {\bibfield  {journal} {\bibinfo  {journal} {Astron. Astrophys.}\ }\textbf
  {\bibinfo {volume} {358}},\ \bibinfo {pages} {521} (\bibinfo {year}
  {2000})},\ \Eprint {http://arxiv.org/abs/astro-ph/0006104}
  {arXiv:astro-ph/0006104 [astro-ph]} \BibitemShut {NoStop}%
\bibitem [{\citenamefont {Lorimer}\ \emph {et~al.}(2006)\citenamefont {Lorimer}
  \emph {et~al.}}]{Lorimer:2006qs}%
  \BibitemOpen
  \bibfield  {author} {\bibinfo {author} {\bibfnamefont {D.~R.}\ \bibnamefont
  {Lorimer}} \emph {et~al.},\ }\href {\doibase
  10.1111/j.1365-2966.2006.10887.x} {\bibfield  {journal} {\bibinfo  {journal}
  {Mon. Not. Roy. Astron. Soc.}\ }\textbf {\bibinfo {volume} {372}},\ \bibinfo
  {pages} {777} (\bibinfo {year} {2006})},\ \Eprint
  {http://arxiv.org/abs/astro-ph/0607640} {arXiv:astro-ph/0607640 [astro-ph]}
  \BibitemShut {NoStop}%
\bibitem [{\citenamefont {Yusifov}\ and\ \citenamefont
  {Kucuk}(2004)}]{Yusifov:2004fr}%
  \BibitemOpen
  \bibfield  {author} {\bibinfo {author} {\bibfnamefont {I.}~\bibnamefont
  {Yusifov}}\ and\ \bibinfo {author} {\bibfnamefont {I.}~\bibnamefont
  {Kucuk}},\ }\href {\doibase 10.1051/0004-6361:20040152} {\bibfield  {journal}
  {\bibinfo  {journal} {Astron. Astrophys.}\ }\textbf {\bibinfo {volume}
  {422}},\ \bibinfo {pages} {545} (\bibinfo {year} {2004})},\ \Eprint
  {http://arxiv.org/abs/astro-ph/0405559} {arXiv:astro-ph/0405559 [astro-ph]}
  \BibitemShut {NoStop}%
\bibitem [{\citenamefont {Inoue}(2011)}]{Inoue:2011bm}%
  \BibitemOpen
  \bibfield  {author} {\bibinfo {author} {\bibfnamefont {Y.}~\bibnamefont
  {Inoue}},\ }\href {\doibase 10.1088/0004-637X/733/1/66} {\bibfield  {journal}
  {\bibinfo  {journal} {Astrophys.J.}\ }\textbf {\bibinfo {volume} {733}},\
  \bibinfo {pages} {66} (\bibinfo {year} {2011})},\ \Eprint
  {http://arxiv.org/abs/1103.3946} {arXiv:1103.3946 [astro-ph.HE]} \BibitemShut
  {NoStop}%
\bibitem [{\citenamefont {Ackermann}\ \emph
  {et~al.}(2012{\natexlab{c}})\citenamefont {Ackermann} \emph
  {et~al.}}]{Ackermann:2012vca}%
  \BibitemOpen
  \bibfield  {author} {\bibinfo {author} {\bibfnamefont {M.}~\bibnamefont
  {Ackermann}} \emph {et~al.} (\bibinfo {collaboration} {Fermi LAT
  Collaboration}),\ }\href {\doibase 10.1088/0004-637X/755/2/164} {\bibfield
  {journal} {\bibinfo  {journal} {Astrophys.J.}\ }\textbf {\bibinfo {volume}
  {755}},\ \bibinfo {pages} {164} (\bibinfo {year} {2012}{\natexlab{c}})},\
  \Eprint {http://arxiv.org/abs/1206.1346} {arXiv:1206.1346 [astro-ph.HE]}
  \BibitemShut {NoStop}%
\bibitem [{\citenamefont {{Di Mauro}}\ \emph {et~al.}(2014)\citenamefont {{Di
  Mauro}}, \citenamefont {{Calore}}, \citenamefont {{Donato}}, \citenamefont
  {{Ajello}},\ and\ \citenamefont {{Latronico}}}]{2014ApJ...780..161D}%
  \BibitemOpen
  \bibfield  {author} {\bibinfo {author} {\bibfnamefont {M.}~\bibnamefont {{Di
  Mauro}}}, \bibinfo {author} {\bibfnamefont {F.}~\bibnamefont {{Calore}}},
  \bibinfo {author} {\bibfnamefont {F.}~\bibnamefont {{Donato}}}, \bibinfo
  {author} {\bibfnamefont {M.}~\bibnamefont {{Ajello}}}, \ and\ \bibinfo
  {author} {\bibfnamefont {L.}~\bibnamefont {{Latronico}}},\ }\href {\doibase
  10.1088/0004-637X/780/2/161} {\bibfield  {journal} {\bibinfo  {journal}
  {Astrophys.J.}\ }\textbf {\bibinfo {volume} {780}},\ \bibinfo {eid} {161}
  (\bibinfo {year} {2014})},\ \Eprint {http://arxiv.org/abs/1304.0908}
  {arXiv:1304.0908 [astro-ph.HE]} \BibitemShut {NoStop}%
\bibitem [{\citenamefont {{Deng}}\ \emph {et~al.}(2011)\citenamefont {{Deng}},
  \citenamefont {{Chen}},\ and\ \citenamefont {{Jiang}}}]{2011MNRAS.417..453D}%
  \BibitemOpen
  \bibfield  {author} {\bibinfo {author} {\bibfnamefont {X.-F.}\ \bibnamefont
  {{Deng}}}, \bibinfo {author} {\bibfnamefont {Y.-Q.}\ \bibnamefont {{Chen}}},
  \ and\ \bibinfo {author} {\bibfnamefont {P.}~\bibnamefont {{Jiang}}},\ }\href
  {\doibase 10.1111/j.1365-2966.2011.19277.x} {\bibfield  {journal} {\bibinfo
  {journal} {Mon.Not.Roy.Astron.Soc.}\ }\textbf {\bibinfo {volume} {417}},\
  \bibinfo {pages} {453} (\bibinfo {year} {2011})}\BibitemShut {NoStop}%
\bibitem [{\citenamefont {{Hodge}}\ \emph {et~al.}(2009)\citenamefont
  {{Hodge}}, \citenamefont {{Zeimann}}, \citenamefont {{Becker}},\ and\
  \citenamefont {{White}}}]{2009AJ....138..900H}%
  \BibitemOpen
  \bibfield  {author} {\bibinfo {author} {\bibfnamefont {J.~A.}\ \bibnamefont
  {{Hodge}}}, \bibinfo {author} {\bibfnamefont {G.~R.}\ \bibnamefont
  {{Zeimann}}}, \bibinfo {author} {\bibfnamefont {R.~H.}\ \bibnamefont
  {{Becker}}}, \ and\ \bibinfo {author} {\bibfnamefont {R.~L.}\ \bibnamefont
  {{White}}},\ }\href {\doibase 10.1088/0004-6256/138/3/900} {\bibfield
  {journal} {\bibinfo  {journal} {Astron.J.}\ }\textbf {\bibinfo {volume}
  {138}},\ \bibinfo {pages} {900} (\bibinfo {year} {2009})},\ \Eprint
  {http://arxiv.org/abs/0907.1081} {arXiv:0907.1081} \BibitemShut {NoStop}%
\bibitem [{\citenamefont {Goodenough}\ and\ \citenamefont
  {Hooper}(2009)}]{Goodenough:2009gk}%
  \BibitemOpen
  \bibfield  {author} {\bibinfo {author} {\bibfnamefont {L.}~\bibnamefont
  {Goodenough}}\ and\ \bibinfo {author} {\bibfnamefont {D.}~\bibnamefont
  {Hooper}},\ }\href@noop {} {\  (\bibinfo {year} {2009})},\ \Eprint
  {http://arxiv.org/abs/0910.2998} {arXiv:0910.2998 [hep-ph]} \BibitemShut
  {NoStop}%
\bibitem [{\citenamefont {Hooper}\ and\ \citenamefont
  {Goodenough}(2011)}]{Hooper:2010mq}%
  \BibitemOpen
  \bibfield  {author} {\bibinfo {author} {\bibfnamefont {D.}~\bibnamefont
  {Hooper}}\ and\ \bibinfo {author} {\bibfnamefont {L.}~\bibnamefont
  {Goodenough}},\ }\href {\doibase 10.1016/j.physletb.2011.02.029} {\bibfield
  {journal} {\bibinfo  {journal} {Phys.Lett.}\ }\textbf {\bibinfo {volume}
  {B697}},\ \bibinfo {pages} {412} (\bibinfo {year} {2011})},\ \Eprint
  {http://arxiv.org/abs/1010.2752} {arXiv:1010.2752 [hep-ph]} \BibitemShut
  {NoStop}%
\bibitem [{\citenamefont {Hooper}\ and\ \citenamefont
  {Linden}(2011)}]{Hooper:2011ti}%
  \BibitemOpen
  \bibfield  {author} {\bibinfo {author} {\bibfnamefont {D.}~\bibnamefont
  {Hooper}}\ and\ \bibinfo {author} {\bibfnamefont {T.}~\bibnamefont
  {Linden}},\ }\href {\doibase 10.1103/PhysRevD.84.123005} {\bibfield
  {journal} {\bibinfo  {journal} {Phys.Rev.}\ }\textbf {\bibinfo {volume}
  {D84}},\ \bibinfo {pages} {123005} (\bibinfo {year} {2011})},\ \Eprint
  {http://arxiv.org/abs/1110.0006} {arXiv:1110.0006 [astro-ph.HE]} \BibitemShut
  {NoStop}%
\bibitem [{\citenamefont {Boyarsky}\ \emph {et~al.}(2011)\citenamefont
  {Boyarsky}, \citenamefont {Malyshev},\ and\ \citenamefont
  {Ruchayskiy}}]{Boyarsky:2010dr}%
  \BibitemOpen
  \bibfield  {author} {\bibinfo {author} {\bibfnamefont {A.}~\bibnamefont
  {Boyarsky}}, \bibinfo {author} {\bibfnamefont {D.}~\bibnamefont {Malyshev}},
  \ and\ \bibinfo {author} {\bibfnamefont {O.}~\bibnamefont {Ruchayskiy}},\
  }\href {\doibase 10.1016/j.physletb.2011.10.014} {\bibfield  {journal}
  {\bibinfo  {journal} {Phys.Lett.}\ }\textbf {\bibinfo {volume} {B705}},\
  \bibinfo {pages} {165} (\bibinfo {year} {2011})},\ \Eprint
  {http://arxiv.org/abs/1012.5839} {arXiv:1012.5839 [hep-ph]} \BibitemShut
  {NoStop}%
\bibitem [{\citenamefont {Gordon}\ and\ \citenamefont
  {Macias}(2013)}]{Gordon:2013vta}%
  \BibitemOpen
  \bibfield  {author} {\bibinfo {author} {\bibfnamefont {C.}~\bibnamefont
  {Gordon}}\ and\ \bibinfo {author} {\bibfnamefont {O.}~\bibnamefont
  {Macias}},\ }\href {\doibase 10.1103/PhysRevD.88.083521} {\bibfield
  {journal} {\bibinfo  {journal} {Phys.Rev.}\ }\textbf {\bibinfo {volume}
  {D88}},\ \bibinfo {pages} {083521} (\bibinfo {year} {2013})},\ \Eprint
  {http://arxiv.org/abs/1306.5725} {arXiv:1306.5725 [astro-ph.HE]} \BibitemShut
  {NoStop}%
\bibitem [{\citenamefont {Daylan}\ \emph {et~al.}(2014)\citenamefont {Daylan},
  \citenamefont {Finkbeiner}, \citenamefont {Hooper}, \citenamefont {Linden},
  \citenamefont {Portillo} \emph {et~al.}}]{Daylan:2014rsa}%
  \BibitemOpen
  \bibfield  {author} {\bibinfo {author} {\bibfnamefont {T.}~\bibnamefont
  {Daylan}}, \bibinfo {author} {\bibfnamefont {D.~P.}\ \bibnamefont
  {Finkbeiner}}, \bibinfo {author} {\bibfnamefont {D.}~\bibnamefont {Hooper}},
  \bibinfo {author} {\bibfnamefont {T.}~\bibnamefont {Linden}}, \bibinfo
  {author} {\bibfnamefont {S.~K.~N.}\ \bibnamefont {Portillo}},  \emph
  {et~al.},\ }\href@noop {} {\  (\bibinfo {year} {2014})},\ \Eprint
  {http://arxiv.org/abs/1402.6703} {arXiv:1402.6703 [astro-ph.HE]} \BibitemShut
  {NoStop}%
\bibitem [{\citenamefont {Abazajian}\ \emph {et~al.}(2015)\citenamefont
  {Abazajian}, \citenamefont {Canac}, \citenamefont {Horiuchi}, \citenamefont
  {Kaplinghat},\ and\ \citenamefont {Kwa}}]{Abazajian:2014hsa}%
  \BibitemOpen
  \bibfield  {author} {\bibinfo {author} {\bibfnamefont {K.~N.}\ \bibnamefont
  {Abazajian}}, \bibinfo {author} {\bibfnamefont {N.}~\bibnamefont {Canac}},
  \bibinfo {author} {\bibfnamefont {S.}~\bibnamefont {Horiuchi}}, \bibinfo
  {author} {\bibfnamefont {M.}~\bibnamefont {Kaplinghat}}, \ and\ \bibinfo
  {author} {\bibfnamefont {A.}~\bibnamefont {Kwa}},\ }\href {\doibase
  10.1088/1475-7516/2015/07/013} {\bibfield  {journal} {\bibinfo  {journal}
  {JCAP}\ }\textbf {\bibinfo {volume} {1507}},\ \bibinfo {pages} {013}
  (\bibinfo {year} {2015})},\ \Eprint {http://arxiv.org/abs/1410.6168}
  {arXiv:1410.6168 [astro-ph.HE]} \BibitemShut {NoStop}%
\bibitem [{\citenamefont {{Regis}}\ \emph {et~al.}(2015)\citenamefont
  {{Regis}}, \citenamefont {{Xia}}, \citenamefont {{Cuoco}}, \citenamefont
  {{Branchini}}, \citenamefont {{Fornengo}},\ and\ \citenamefont
  {{Viel}}}]{2015PhRvL.114x1301R}%
  \BibitemOpen
  \bibfield  {author} {\bibinfo {author} {\bibfnamefont {M.}~\bibnamefont
  {{Regis}}}, \bibinfo {author} {\bibfnamefont {J.-Q.}\ \bibnamefont {{Xia}}},
  \bibinfo {author} {\bibfnamefont {A.}~\bibnamefont {{Cuoco}}}, \bibinfo
  {author} {\bibfnamefont {E.}~\bibnamefont {{Branchini}}}, \bibinfo {author}
  {\bibfnamefont {N.}~\bibnamefont {{Fornengo}}}, \ and\ \bibinfo {author}
  {\bibfnamefont {M.}~\bibnamefont {{Viel}}},\ }\href {\doibase
  10.1103/PhysRevLett.114.241301} {\bibfield  {journal} {\bibinfo  {journal}
  {Physical Review Letters}\ }\textbf {\bibinfo {volume} {114}},\ \bibinfo
  {eid} {241301} (\bibinfo {year} {2015})},\ \Eprint
  {http://arxiv.org/abs/1503.05922} {arXiv:1503.05922} \BibitemShut {NoStop}%
\bibitem [{\citenamefont {Ackermann}\ \emph
  {et~al.}(2015{\natexlab{b}})\citenamefont {Ackermann} \emph
  {et~al.}}]{Ackermann:2015tah}%
  \BibitemOpen
  \bibfield  {author} {\bibinfo {author} {\bibfnamefont {M.}~\bibnamefont
  {Ackermann}} \emph {et~al.} (\bibinfo {collaboration} {Fermi-LAT}),\ }\href
  {\doibase 10.1088/1475-7516/2015/09/008} {\bibfield  {journal} {\bibinfo
  {journal} {JCAP}\ }\textbf {\bibinfo {volume} {1509}},\ \bibinfo {pages}
  {008} (\bibinfo {year} {2015}{\natexlab{b}})},\ \Eprint
  {http://arxiv.org/abs/1501.05464} {arXiv:1501.05464 [astro-ph.CO]}
  \BibitemShut {NoStop}%
\bibitem [{\citenamefont {{Kuhlen}}\ \emph {et~al.}(2008)\citenamefont
  {{Kuhlen}}, \citenamefont {{Diemand}},\ and\ \citenamefont
  {{Madau}}}]{2008ApJ...686..262K}%
  \BibitemOpen
  \bibfield  {author} {\bibinfo {author} {\bibfnamefont {M.}~\bibnamefont
  {{Kuhlen}}}, \bibinfo {author} {\bibfnamefont {J.}~\bibnamefont {{Diemand}}},
  \ and\ \bibinfo {author} {\bibfnamefont {P.}~\bibnamefont {{Madau}}},\ }\href
  {\doibase 10.1086/590337} {\bibfield  {journal} {\bibinfo  {journal}
  {Astrophys.J.}\ }\textbf {\bibinfo {volume} {686}},\ \bibinfo {pages} {262}
  (\bibinfo {year} {2008})},\ \Eprint {http://arxiv.org/abs/0805.4416}
  {arXiv:0805.4416} \BibitemShut {NoStop}%
\bibitem [{\citenamefont {{Kamionkowski}}\ \emph {et~al.}(2010)\citenamefont
  {{Kamionkowski}}, \citenamefont {{Koushiappas}},\ and\ \citenamefont
  {{Kuhlen}}}]{2010PhRvD..81d3532K}%
  \BibitemOpen
  \bibfield  {author} {\bibinfo {author} {\bibfnamefont {M.}~\bibnamefont
  {{Kamionkowski}}}, \bibinfo {author} {\bibfnamefont {S.~M.}\ \bibnamefont
  {{Koushiappas}}}, \ and\ \bibinfo {author} {\bibfnamefont {M.}~\bibnamefont
  {{Kuhlen}}},\ }\href {\doibase 10.1103/PhysRevD.81.043532} {\bibfield
  {journal} {\bibinfo  {journal} {Phys.Rev.}\ ,\ \bibinfo {eid} {043532}}
  (\bibinfo {year} {2010})},\ \Eprint {http://arxiv.org/abs/1001.3144}
  {arXiv:1001.3144 [astro-ph.GA]} \BibitemShut {NoStop}%
\bibitem [{\citenamefont {{Zavala}}\ and\ \citenamefont
  {{Afshordi}}(2014)}]{2014MNRAS.441.1329Z}%
  \BibitemOpen
  \bibfield  {author} {\bibinfo {author} {\bibfnamefont {J.}~\bibnamefont
  {{Zavala}}}\ and\ \bibinfo {author} {\bibfnamefont {N.}~\bibnamefont
  {{Afshordi}}},\ }\href {\doibase 10.1093/mnras/stu506} {\bibfield  {journal}
  {\bibinfo  {journal} {Mon.Not.Roy.Astron.Soc.}\ }\textbf {\bibinfo {volume}
  {441}},\ \bibinfo {pages} {1329} (\bibinfo {year} {2014})},\ \Eprint
  {http://arxiv.org/abs/1311.3296} {arXiv:1311.3296} \BibitemShut {NoStop}%
\bibitem [{\citenamefont {{Okabe}}\ \emph {et~al.}(2010)\citenamefont
  {{Okabe}}, \citenamefont {{Okura}},\ and\ \citenamefont
  {{Futamase}}}]{2010ApJ...713..291O}%
  \BibitemOpen
  \bibfield  {author} {\bibinfo {author} {\bibfnamefont {N.}~\bibnamefont
  {{Okabe}}}, \bibinfo {author} {\bibfnamefont {Y.}~\bibnamefont {{Okura}}}, \
  and\ \bibinfo {author} {\bibfnamefont {T.}~\bibnamefont {{Futamase}}},\
  }\href {\doibase 10.1088/0004-637X/713/1/291} {\bibfield  {journal} {\bibinfo
   {journal} {Astron.J.}\ }\textbf {\bibinfo {volume} {713}},\ \bibinfo {pages}
  {291} (\bibinfo {year} {2010})},\ \Eprint {http://arxiv.org/abs/1001.2402}
  {arXiv:1001.2402} \BibitemShut {NoStop}%
\bibitem [{\citenamefont {{Li}}\ \emph {et~al.}(2014)\citenamefont {{Li}},
  \citenamefont {{Shan}}, \citenamefont {{Mo}}, \citenamefont {{Kneib}},
  \citenamefont {{Yang}}, \citenamefont {{Luo}}, \citenamefont {{van den
  Bosch}}, \citenamefont {{Erben}}, \citenamefont {{Moraes}},\ and\
  \citenamefont {{Makler}}}]{2014MNRAS.438.2864L}%
  \BibitemOpen
  \bibfield  {author} {\bibinfo {author} {\bibfnamefont {R.}~\bibnamefont
  {{Li}}}, \bibinfo {author} {\bibfnamefont {H.}~\bibnamefont {{Shan}}},
  \bibinfo {author} {\bibfnamefont {H.}~\bibnamefont {{Mo}}}, \bibinfo {author}
  {\bibfnamefont {J.-P.}\ \bibnamefont {{Kneib}}}, \bibinfo {author}
  {\bibfnamefont {X.}~\bibnamefont {{Yang}}}, \bibinfo {author} {\bibfnamefont
  {W.}~\bibnamefont {{Luo}}}, \bibinfo {author} {\bibfnamefont {F.~C.}\
  \bibnamefont {{van den Bosch}}}, \bibinfo {author} {\bibfnamefont
  {T.}~\bibnamefont {{Erben}}}, \bibinfo {author} {\bibfnamefont
  {B.}~\bibnamefont {{Moraes}}}, \ and\ \bibinfo {author} {\bibfnamefont
  {M.}~\bibnamefont {{Makler}}},\ }\href {\doibase 10.1093/mnras/stt2395}
  {\bibfield  {journal} {\bibinfo  {journal} {Mon.Not.Roy.Astron.Soc.}\
  }\textbf {\bibinfo {volume} {438}},\ \bibinfo {pages} {2864} (\bibinfo {year}
  {2014})},\ \Eprint {http://arxiv.org/abs/1311.6523} {arXiv:1311.6523
  [astro-ph.CO]} \BibitemShut {NoStop}%
\bibitem [{\citenamefont {{Shirasaki}}(2015)}]{2015ApJ...799..188S}%
  \BibitemOpen
  \bibfield  {author} {\bibinfo {author} {\bibfnamefont {M.}~\bibnamefont
  {{Shirasaki}}},\ }\href {\doibase 10.1088/0004-637X/799/2/188} {\bibfield
  {journal} {\bibinfo  {journal} {Astrophys.J.}\ }\textbf {\bibinfo {volume}
  {799}},\ \bibinfo {eid} {188} (\bibinfo {year} {2015})},\ \Eprint
  {http://arxiv.org/abs/1407.5350} {arXiv:1407.5350} \BibitemShut {NoStop}%
\end{thebibliography}%
\end{document}